\RequirePackage{snapshot}

\documentclass[10pt]{llncs}

\usepackage[backgroundcolor=orange!50!white]{todonotes}

\usepackage{cesar-llncs}
\usepackage{amsmath}
\usepackage{amsfonts, amssymb}
\usepackage{latexsym, stmaryrd, xspace}
\usepackage{ltlfonts}

\usepackage{graphicx}
\usepackage{float}
\usepackage{xspace}
\usepackage{listings}
\usepackage{algorithm}
\usepackage{algpseudocode}
\usepackage{ale-defs}
\usepackage{cite}
\usepackage{bussproofs}

\usepackage{array}

\usepackage{mdwmath}
\usepackage{mdwtab}
\usepackage{eqparbox}

\usepackage{url}
\usepackage{subfig}
\usepackage{paralist}
\usepackage{afterpage}
\usepackage{wrapfig}

\usepackage{calc}

  \algsetblockdefx[MyLoop]{LOOP}{ENDLOOP}{65535}{3em}{\hspace{1.5em}\textbf{loop}}{\hspace{1em}\textbf{end loop}}
  
\newcommand{\elabel}[1]{ 
  \renewcommand{\@currentlabel}{\S\PrintEntryNumber}
  \label{#1}
}

\newcommand{\PrintEntryTitle}{}
\newcommand{\PrintEntryNumber}{}
\newcommand{\PrintEntryDate}{}
\newcommand{\PrintEntryModified}{}

\newcommand{\entrydate}[1]{
  \ifthenelse{\equal{#1}{}}
  {\renewcommand{\PrintEntryDate}{}}
  {\renewcommand{\PrintEntryDate}{
    \linebreak\hspace*{\fill}\textit{Date: #1}}}
}
\newcommand{\entrymodified}[1]{
  \ifthenelse{\equal{#1}{}}
  {\renewcommand{\PrintEntryModified}{}}
  {\renewcommand{\PrintEntryModified}{
      \linebreak\hspace*{\fill}\textit{Last Modified: #1}}}
}

\newcounter{EntryCount}
  
\newcommand\Def[2][]{
  \def\temp{#1}
  \ifx\temp\empty%
    \def\temp{#2}%
  \fi%
{\em#2\index{\temp}}}

\newcommand{\onlyWithComplement}[1]{}

\providecommand*{\Always}{} 
\renewcommand{\Always}{\LTLsquare}

\newcommand{\SEQsym}{{\scriptstyle{\cdot}}}

\newcommand{\OR}{\mathrel{+}}
\newcommand{\AND}{\mathrel{|}} 
\newcommand{\SEQ}{\mathrel{\cdot}}

\newcommand{\Eqs}{\mathcal{E}}

\renewcommand{\SEQsym}{\ensuremath{;}}
\renewcommand{\SEQ}{\mathrel{\SEQsym}}

\input{macros/rules}

\newcommand{\mutex}{\ensuremath{\mathsf{mutex}}\xspace}
\newcommand{\setmutex}{\ensuremath{\mathsf{mutexS}}\xspace}

\newcommand{\Active}{\ensuremath{\mathsf{active}}\xspace}

\newcommand{\Critical}{\ensuremath{\mathsf{critical}}\xspace}

\newcommand{\minticket}{\ensuremath{\mathsf{minticket}}\xspace}
\newcommand{\activelow}{\ensuremath{\mathsf{activelow}}\xspace}
\newcommand{\notsame}{\ensuremath{\mathsf{notsame}}\xspace}

\newcommand{\setminticket}{\ensuremath{\mathsf{minticketS}}\xspace}
\newcommand{\setactivelow}{\ensuremath{\mathsf{activelowS}}\xspace}
\newcommand{\setnotsame}{\ensuremath{\mathsf{notsameS}}\xspace}

\newcommand{\supports}{\mathrel{\rhd}}

\newcommand{\INV}{\textsc{inv}\xspace}
\newcommand{\bINV}{\textsc{b-inv}\xspace}

\newcommand{\pINV}{\textsc{p-inv}\xspace}
\newcommand{\spINV}{\textsc{sp-inv}\xspace}
\newcommand{\gINV}{\textsc{g-inv}\xspace}

\newcommand{\sArray}{\textsf{array}\xspace}
\newcommand{\Arr}{\textit{Arr}} 

\newcommand{\Premise}[1]{\ensuremath{\mathsf{#1}}\xspace}

\newcommand{\bINVRule}{
  \framebox{
    \begin{minipage}[b]{17.3em}
      To show that $\Sys$ satisfies $\Always \varphi$:
       \[ \begin{array}{r@{\hspace{2em}}r@{\;}c@{\;}l@{}l}
         \Premise{B1.} & \Theta &\Into& \varphi & \\
         \Premise{B2.} &  \varphi \And \tau &\Into& \varphi'  & \;\;\;\;\text{for all $\tau$}\\
         \phantom{\Premise{I3}} \\ \cline{1-5}
         & \multicolumn{3}{c}{\hspace{1em}\Always \varphi} &
       \end{array}
       \]
    \end{minipage}
  }
}

\newcommand{\INVRule}{
  \framebox{
    \begin{minipage}[b]{17.3em}
      To show that $\Sys$ satisfies $\Always p$, find $q$ with:
       \[ \begin{array}{r@{\hspace{2em}}r@{\;}c@{\;}l@{}l}
         \Premise{I1}. & \Theta &\Into& q & \\
         \Premise{I2}. &  q \And \tau &\Into& q'  & \;\;\;\;\text{for all $\tau$}\\
         \Premise{I3}. & q &\Into& \varphi \\ \cline{1-5}
         & \multicolumn{3}{c}{\hspace{1em}\Always \varphi} &
       \end{array}
       \]
    \end{minipage}
  }
}

\newcommand{\pINVRule}{
  \framebox{
    \begin{minipage}[b]{30em}
      To show that $\Sys$ satisfies $\Always\varphi(\overline{k})$, with $\overline{k}=\Var(\varphi)$:
      \hspace{-1em}
      \[ \begin{array}{r@{\;\;}lr@{\;}@{\;}cl@{\hspace{1em}}l}
         \Premise{P1}. & & \Theta(\overline{k}) &\Into& \varphi & \\
         \Premise{P2}. & & \varphi \And \tau^{(i)} &\Into& \varphi'  & \text{forall $\tau$ and all $i\in \overline{k}$}\\
         \Premise{P3}. & &\varphi \And \big(\bigwedge\limits_{x\in\overline{k}} j\neq x \And \tau^{(j)} &\Into& \varphi' \big)& \text{forall $\tau$ and one fresh $j\notin\overline{k}$}\\ \hline
         & \multicolumn{4}{c}{\hspace{3em} \Always \varphi} &
       \end{array}
       \]
    \end{minipage}
  }
}

\newcommand{\pINVThree}{
  \framebox{
    \begin{minipage}[b]{30em}
      \[
      \begin{array}{r@{\;\;}lr@{\;}@{\;}cl@{\hspace{1em}}l}
        \;\;\Premise{P3'}. & \varphi \supports & \big(\bigwedge\limits_{x\in\overline{k}} j\neq x \And \tau^{(j)} &\Into& \varphi' \big)& \text{forall $\tau$ and one fresh $j\notin\overline{k}$}\;\;
      \end{array}\]
  \end{minipage}
}}

\newcommand{\spINVRule}{
  \centering
  \framebox{
    \begin{minipage}[b]{35em}
      To show that $\Sys$ satisfies $\Always\varphi(\overline{j})$. Find $\psi(\overline{k})$ with:
      \[ \begin{array}{r@{\hspace{2em}}rcl@{\hspace{2em}}l}
        \Premise{S1.} \;\Premise{S2.}\; \Premise{S3.} & \multicolumn{1}{l}{\Premise{U1.}\; \Premise{U2.}\; \Premise{U3.}} & & &\\
  \Premise{S4.} & \bigwedge\limits_{\sigma\in S} \psi\;\sigma \And \varphi \And \tau^{(i)} &\Into& \varphi' &  \text{for all $\tau$}, i\in\overline{j}, \text{and}\\
  & & & & \;\;\;\;\;S=\Arr(\overline{k},\overline{k}\cup\overline{j})\\
  \Premise{S5.} & \bigwedge\limits_{x\in\overline{j}} i\neq x \And \bigwedge\limits_{\sigma\in S} \psi\;\sigma \And \varphi  \And
  \tau^{(i)} &\Into& \varphi' & \text{for all $\tau$, $i\notin\overline{j}$},\text{and}\\ 
  & & & &\;\;\;\;\;S=\Arr(\overline{k},\overline{k}\cup\overline{j}\cup\{i\})\\
        \hline &
        \multicolumn{3}{c}{\Always\varphi} &
       \end{array}
       \]
    \end{minipage}
  }
}

\newcommand{\strgINVRule}{
  \centering
  \framebox{
    \begin{minipage}[b]{34em}
      To show that $\Sys$ satisfies $\Always\varphi(\overline{k})$. Find $\psi$ with:
      \[ \begin{array}{r@{\hspace{1em}}r@{\hspace{1em}}rcl@{\hspace{2em}}l}
        \Premise{S0.} & &\multicolumn{3}{c}{\hspace{2em}\Always\psi} & \\
        \Premise{S1.} & & \Theta & \Into & \varphi & \\
  \Premise{S2.} & \psi,\varphi \supports &  \tau^{(i)} &\Into& \varphi' &  \text{forall $\tau$ and all $i\in \overline{k}$}\\
  \Premise{S3.} & \psi,\varphi\supports & \bigwedge\limits_{x\in\overline{k}} j\neq x   \And
  \tau^{(j)} &\Into& \varphi' & \text{forall $\tau$ and one fresh $j\notin\overline{k}$}\\ \hline 
  & &\multicolumn{3}{c}{\hspace{2em}\Always\varphi} &
       \end{array}
       \]
    \end{minipage}
  }
}

\newcommand{\GeneralGraphINVRule}{
  \centering
  \framebox{
    \begin{minipage}[b]{34em}
      To show that $\Sys$ satisfies $\Always\varphi$ find a proof graph $(V,E)$ with $\varphi\in V$ such that:
      \[ \begin{array}{l@{\hspace{1em}}r@{\hspace{1em}}rcl@{\hspace{2em}}l}
        \Premise{G1.} & & \Theta & \Into & \psi & \text{forall $\psi\in V$}\\
\Premise{G2.} & \Phi,\psi \supports & \tau^{(k)} &\Into& \psi' &  \text{forall $\psi\in{V}$, forall $\tau$,}\\ 
 & & & & &\;\;\;\;\;\;\;\text{and all $k\in \Var(\psi)$,}\\
 & & & & &\;\;\;\;\;\;\;\text{and $\Phi=\{\psi_i\;|\;(\psi_i,\psi)\in E\}$}\\
         \Premise{G3.} & \Phi,\psi \supports & 
 \bigwedge\limits_{x\in v} k\neq x   \And
   \tau^{(k)} &\Into& \varphi' &  \text{forall $\psi\in V$, forall $\tau$,}\\
         & & &  & &\;\;\;\;\;\;\;\text{one fresh $k\notin v=\Var(\psi)$,}\\
         & & &  & &\;\;\;\;\;\;\;\text{and $\Phi=\{\psi_i\;|\;(\psi_i,\psi)\in E\}$}\\ \hline
          & &\multicolumn{3}{c}{\hspace{2em}\Always\varphi} &
       \end{array}
       \]
    \end{minipage}
  }
}

\newcommand{\sInt}{\textsf{int}\xspace}
\newcommand{\sBool}{\textsf{bool}\xspace}
\newcommand{\setTh}{\textsf{settid}\xspace}
\newcommand{\sTID}{\textsf{tid}\xspace}
\newcommand{\Loc}{\textit{Loc}\xspace}
\newcommand{\Vglobal}{\ensuremath{V_{\textit{global}}}\xspace}

\newcommand{\Vlocal}{\ensuremath{V_{\textit{local}}}\xspace}

\newcommand{\deltaij}{\ensuremath{\pi_{ij}}\xspace}
\newcommand{\deltaji}{\ensuremath{\pi_{ji}}\xspace}

\newcommand{\M}{\mathcal{M}}
\newcommand{\deltaijM}{\ensuremath{\pi_{ij}^{\M}}\xspace}
\newcommand{\deltajiM}{\ensuremath{\pi_{ji}^{\M}}\xspace}
\newcommand{\deltaijE}{\ensuremath{\pi_{ij}^{E}}\xspace}
\newcommand{\deltaijS}{\ensuremath{\pi_{ij}^{S}}\xspace}
\newcommand{\deltaijT}{\ensuremath{\pi_{ij}^{T}}\xspace}
\newcommand{\deltaijP}{\ensuremath{\pi_{ij}^{P}}\xspace}

\newcommand{\swap}{\ensuremath{\textit{swap}}}

\newcommand{\sigTID}{\ensuremath{\Sigma_{\sTID}}\xspace}
\newcommand{\TTID}{\ensuremath{T_{\sTID}}\xspace}

\newcommand{\TArray}{\ensuremath{T_{\textsf{A}}}\xspace}

\newcommand{\sProg}{\textsf{prog}}
\newcommand{\sigProg}{\ensuremath{\Sigma_{\sProg}}\xspace}
\newcommand{\TProg}{\ensuremath{T_{\sProg}}\xspace}

\newcommand{\pc}{\ensuremath{\textit{pc}}}

\newcommand{\ThetaGlobal}{\Theta_{g}}
\newcommand{\ThetaLocal}{\Theta_{l}}
\usepackage{hhline}

\newcolumntype{W}{>{\raggedleft\arraybackslash}m{4.2em}<{}}
\newcolumntype{T}{>{\raggedleft\arraybackslash}m{2em}<{}}
\newcommand{\MulTwo}[1]{\multicolumn{2}{c||}{#1}}

\newcommand{\ColC}[1]{\multicolumn{1}{c|}{#1}}
\newcommand{\ColCC}[1]{\multicolumn{1}{c||}{#1}}

\newcommand{\ResultsTable}{
	\begin{tabular}{ |r||T|T||T|T||W|W|W|W|} \cline{2-9}
		\ColC{} & \MulTwo{formula info} & \MulTwo{\#solved vc} &
		\ColC{Full supp} & \ColC{Supp} & \ColC{Offend} & \ColC{Tactics} \\
		\cline{2-5}
		\ColC{}& \ColC{index} & \ColCC{\#vc} & \ColC{pos} & \ColCC{dp} &
		\ColC{time(s)} & \ColC{time(s)} & \ColC{time(s)} & \ColC{time(s)} \\
		\hhline{-*{8}=}
		$\invPreserve$	 & $0$ & $ 61$ & $ 38$ & $23$ & $      TO$ & $     TO$ & $     TO$ & $12.85$ \\ \hline
		$\invOrder$			 & $1$ & $121$ & $ 62$ & $59$ & $  998.35$ & $   7.56$ & $   2.69$ & $ 1.20$ \\ \hline
		$\invLock$			 & $1$ & $121$ & $ 76$ & $45$ & $  778.15$ & $   4.82$ & $   1.44$ & $ 0.50$ \\ \hline
		$\invNext$			 & $1$ & $121$ & $ 60$ & $61$ & $      TO$ & $     TO$ & $  26.58$ & $ 1.76$ \\ \hline
		$\invRegion$		 & $1$ & $121$ & $ 95$ & $26$ & $      TO$ & $     TO$ & $  85.27$ & $25.67$ \\ \hline
		$\invDisjoint$	 & $2$ & $181$ & $177$ & $ 4$ & $  121.74$ & $   1.29$ & $   1.29$ & $ 0.22$ \\ \hline \hline
		$\invFuncSearch$ & $1$ & $208$ & $198$ & $10$ & $      TO$ & $     TO$ & $   6.14$ & $ 4.55$ \\ \hline
		$\invFuncInsert$ & $1$ & $208$ & $200$ & $ 8$ & $      TO$ & $     TO$ & $   2.04$ & $ 0.51$ \\ \hline
		$\invFuncRemove$ & $1$ & $208$ & $200$ & $ 8$ & $      TO$ & $     TO$ & $   2.73$ & $ 1.56$ \\ \hline \hline

		$\invFunSchLinear$& $1$ & $121$ &  $97$ & $24$ & $      TO$ & $     TO$ & $ 82.13$ & $ 4.63$ \\ \hline
		$\invFunSchInsert$& $1$ & $121$ &  $93$ & $28$ & $      TO$ & $     TO$ & $ 80.20$ & $ 5.00$ \\ \hline
		$\invFunSchRemove$& $1$ & $121$ &  $93$ & $28$ & $      TO$ & $     TO$ & $110.84$ & $ 5.49$ \\ \hline \hline
		$\mutex$     		 & $2$ &  $28$ &  $26$ &  $2$ & $    0.32$ & $   0.23$ & $   0.10$ & $ 0.01$ \\ \hline
		$\minticket$  	 & $1$ &  $19$ &  $18$ &  $1$ & $    0.04$ & $   0.04$ & $   0.01$ & $ 0.01$ \\ \hline
		$\notsame$ 			 & $2$ &  $28$ &  $26$ &  $2$ & $    0.13$ & $   0.13$ & $   0.10$ & $ 0.02$ \\ \hline
		$\activelow$ 		 & $1$ &  $19$ &  $17$ &  $2$ & $    0.01$ & $   0.01$ & $   0.01$ & $ 0.01$ \\ \hline \hline
		$\setmutex$      & $2$ &  $28$ &  $26$ &  $2$ & $    0.44$ & $   0.38$ & $   0.14$ & $ 0.04$ \\ \hline
		$\setminticket$  & $1$ &  $19$ &  $18$ &  $1$ & $    0.31$ & $   0.18$ & $   0.08$ & $ 0.01$ \\ \hline
		$\setnotsame$ 	 & $2$ &  $28$ &  $26$ &  $2$ & $    0.14$ & $   0.13$ & $   0.10$ & $ 0.02$ \\ \hline
		$\setactivelow$  & $1$ &  $19$ &  $17$ &  $2$ & $    0.02$ & $   0.02$ & $   0.02$ & $ 0.01$ \\ \hline
	\end{tabular}
}

\newcommand{\lblMainE}[1]{\ensuremath{\pc(#1)=3..7}\xspace}
\newcommand{\lblSearchBody}[1]{\ensuremath{\pc(#1)=8..21}\xspace}
\newcommand{\lblInsertBody}[1]{\ensuremath{\pc(#1)=23..40}\xspace}
\newcommand{\lblRemoveBody}[1]{\ensuremath{\pc(#1)=42..58}\xspace}
\newcommand{\lblInsOwnsPrev}[1]{\ensuremath{\pc(#1)=25..39}\xspace}
\newcommand{\lblInsOwnsCurr}[1]{\ensuremath{\pc(#1)=27..31,34..40}\xspace}
\newcommand{\lblInsAuxBefPrev}[1]{\ensuremath{\pc(#1)=30}\xspace}

\newcommand{\lblInsPrevDef}[1]{\ensuremath{\pc(#1)=24..39}\xspace}
\newcommand{\lblInsPrevLower}[1]{\ensuremath{\pc(#1)=24..37}\xspace}
\newcommand{\lblInsCurrDef}[1]{\ensuremath{\pc(#1)=26..40}\xspace}
\newcommand{\lblInsCurrLower}[1]{\ensuremath{\pc(#1)=28..31}\xspace}
\newcommand{\lblInsMalloc}[1]{\ensuremath{\pc(#1)=33,34}\xspace}

\newcommand{\lblInsInsert}[1]{\ensuremath{\pc(#1)=35..37}\xspace}
\newcommand{\lblRemOwnsPrev}[1]{\ensuremath{\pc(#1)=44..57}\xspace}
\newcommand{\lblRemOwnsCurr}[1]{\ensuremath{\pc(#1)=46..50,53..58}\xspace}
\newcommand{\lblRemAuxBefPrev}[1]{\ensuremath{\pc(#1)=49}\xspace}

\newcommand{\lblRemPrevDef}[1]{\ensuremath{\pc(#1)=43..57}\xspace}
\newcommand{\lblRemCurrDef}[1]{\ensuremath{\pc(#1)=45..55}\xspace}

\newcommand{\lblRemRemove}[1]{\ensuremath{\pc(#1)=54,55}\xspace}

\newcommand{\formulaFullListReduced}{
	\begin{array}{lcl}
		\phiPreserve & \defsym &
			\left\{
				\begin{array}{l>{\hspace{1em}}l>{\hspace{1em}}c}
					\fNull \in \region \And
					\region = \fAddrToSet(\heap, \head) \And
					\head \neq \tail & \And & \text{(L1)} \\
					\heap[\tail].\fNext = \fNull \And
					\tail \neq \fNull \And
					\head \neq \fNull & \And & \text{(L2)} \\
					\heap[\head].\fData = -\infty \And
					\heap[\tail].\fData = +\infty & \And & \text{(L3)} \\
					\elements = \fSetToElemSet (\heap, \region) \And
					\Ordered (\heap, \head, \tail) && \text{(L4)}
				\end{array}
			\right.
	\end{array}
}

\newcommand{\formulaFullLockReduced}{
	\begin{array}{rcl}
		\phiLock(i) & \defsym &
			\left\{
				\begin{array}{l>{\hspace{0.5em}}c>{\hspace{0.5em}}l>{\hspace{1em}}l}
					\lblInsOwnsPrev{i} & \Impl &
						\heap[\local{I}{\prev}(i)].\fLockID = i & \And \\
					\lblInsOwnsCurr{i} & \Impl &
						\heap[\local{I}{\curr}(i)].\fLockID = i & \And \\
					\lblInsAuxBefPrev{i} & \Impl &
						\heap[\local{I}{\aux}(i)].\fLockID = i & \And \\
					\lblRemOwnsPrev{i} & \Impl &
						\heap[\local{R}{\prev}(i)].\fLockID = i & \And \\
					\lblRemOwnsCurr{i} & \Impl &
						\heap[\local{R}{\curr}(i)].\fLockID = i & \And \\
					\lblRemAuxBefPrev{i} & \Impl &
						\heap[\local{R}{\aux}(i)].\fLockID = i
				\end{array}
			\right.
	\end{array}
}

\newcommand{\formulaFullRegionReduced}{
	\begin{array}{rcl}
		\phiRegion(i) & \defsym &
			\left\{
				\begin{array}{l>{\hspace{0.5em}}l>{\hspace{0.5em}}c}
					\{ \head, \tail, \fNull \} \subseteq \region \And
					\tail \neq \fNull \And
					\head \neq \fNull \And
					\head \neq \tail & \And \\
					\lblInsPrevDef{i} \Impl
						\local{I}{\prev}(i) \in \region \And
					\lblInsCurrDef{i} \Impl
						\local{I}{\curr}(i) \in \region & \And \\
					\lblInsMalloc{i} \Impl
						\Not \local{I}{aux}(i) \in \region \And
					\lblInsAuxBefPrev{i} \Impl
						\local{I}{\aux}(i) \in \region & \And \\
					\lblRemPrevDef{i} \Impl
						\big(
							\local{R}{\prev}(i) \cap \{ \tail, \fNull \} = \emptyset \And
							\local{R}{\prev}(i) \in \region
						\big) & \And \\
					\lblRemCurrDef{i} \Impl
						\big(
							\local{R}{\curr}(i) \neq \fNull \And
							\local{R}{\curr}(i) \in \region
						\big) & \And \\
					\lblRemAuxBefPrev{i} \Impl
						\local{R}{\aux}(i) \in \region
				\end{array}
			\right.
	\end{array}
}

\newcommand{\formulaFullDisjoint}{
	\begin{array}{rcl}
		\phiDisjoint(i,j) & \defsym &
				(i \neq j \And \lblInsMalloc{i} \And \lblInsMalloc{j})
					\; \Impl \; \local{I}{\aux}(i) \neq \local{I}{\aux}(j)
	\end{array}
}

\newcommand{\formulaFullOrderReduced}{
	\begin{array}{rcl}
		\phiOrder(i) & \defsym &
			\left\{
				\begin{array}{l>{\hspace{0.5em}}l}
					\heap[\head].\fData = - \infty \And
					\heap[\tail].\fData = + \infty & \And \\
					\lblMainE{i} \Impl
						\local{C}{e}(i) \notin \{ \pm \infty \} \And
					\lblSearchBody{i} \Impl
						\local{S}{e}(i) \notin \{ \pm \infty \} & \And \\
					\lblInsertBody{i} \Impl
						\local{I}{e}(i) \notin \{ \pm \infty \} \And
					\lblRemoveBody{i} \Impl
						\local{R}{e}(i) \notin \{ \pm \infty \} & \And \\
					\lblInsCurrDef{i} \Impl
						\heap[\local{I}{\curr}(i)].\fData \leq
						+\infty & \And \\
					\lblInsPrevDef{i} \Impl
						\heap[\local{I}{\prev}(i)].\fData \leq
						+\infty & \And \\
					\lblInsCurrLower{i} \Impl
						\heap[\local{I}{\curr}(i)].\fData < \local{I}{e}(i) & \And \\
					\lblInsPrevLower{i} \Impl
						\heap[\local{I}{\prev}(i)].\fData < \local{I}{e}(i) & \And \\
					\lblInsInsert{i} \Impl
						\local{I}{e}(i) < \heap[\local{I}{\curr}(i)].\fData & \And \\
					\lblInsMalloc{i} \Impl
						\heap[\local{I}{\aux}(i)].\fData = \local{I}{e}(i) & \And \\
					\lblRemRemove{i} \Impl
						\heap[\local{R}{\curr}(i)].\fData = \local{R}{e}(i)
				\end{array}
			\right.
	\end{array}
}

\newcommand{\partialmap}{\mathrel{\rightharpoonup}}

\begin{document}

\title{Parametrized Invariance for\\ Infinite State Processes}

\author{Alejandro S\'anchez\inst{1}\and C\'esar S\'anchez\inst{1,2}}
\institute{IMDEA Software Institute, Madrid, Spain
  \and
  Institute for Information Security, CSIC, Spain\\
  \email{\{alejandro.sanchez,cesar.sanchez\}@imdea.org}
}

\mainmatter
\pagestyle{headings}

\maketitle
\thispagestyle{empty}

\begin{abstract}


  We study the uniform verification problem for infinite state
  processes, which consists of proving that the parallel composition
  of an arbitrary number of processes satisfies a temporal
  property. Our practical motivation is to build a general framework
  for the temporal verification of concurrent datatypes.

  The contribution of this paper is a general method for the
  verification of safety properties of parametrized programs that
  manipulate complex local and global data, including mutable state in
  the heap. This method is based on the separation between two
  concerns: (1) the interaction between executing threads---handled by
  novel parametrized invariance rules---,and the data being
  manipulated---handled by specialized decision procedures.  The proof
  rules discharge automatically a finite collection of verification
  conditions (VCs), the number depending only on the size of the
	program description and the specification, but not on the number of
  processes in any given instance or on the kind of data
  manipulated. Moreover, all VCs are quantifier free, which eases the
  development of decision procedures for complex data-types on top of
  off-the-shelf SMT solvers.

  We discuss the practical verification (of shape and also functional
  correctness properties) of a concurrent list implementation based on
  the method presented in this paper.  Our tool also all VCs using a
  decision procedure for a theory of list layouts in the heap built on
  top of state-of-the-art SMT solvers.
\end{abstract}

\newcommand{\myparagraph}[1]{\paragraph*{\textup{\textbf{#1.}}}}

\section{Introduction}
\label{sec:introduction}

In this paper we present a general method to verify concurrent
software which is run by an arbitrary number of threads that
manipulate complex data, including infinite local and shared state.
Our solution consists of a method that cleanly \emph{separates} two
concerns:
\begin{inparaenum}[(1)]
\item the data, handled by specialized decision procedures; and
\item the concurrent thread interactions which is handled by novel
  proof rules, that we call \emph{parametrized invariance}.
\end{inparaenum}
The method of parametrized invariance tackles, for safety properties,
the \emph{uniform verification} problem for \emph{parametrized
	systems with infinite state processes}:

\noindent\fbox{
  \parbox{\textwidth}{Given a parametrized system
    $\Sys[N]:P(1) \parallel P(2) \parallel \ldots \parallel P(N)$ and
    a property $\varphi$ establish whether $\Sys[N]\vDash \varphi$ for
    all instances $N\geq{}1$.}}

In this paper we restrict to safety properties. 
%
%
Our method is a generalization of the inductive invariance rule for
temporal deductive verification~\cite{manna95temporal}, in which each
verification condition corresponds to a small-step (a single
transition) in the execution of a system.  For non-parametrized
systems, there is always a finite number of transitions, so one can
generate one VC per transition. However, in parametrized systems, the
number of transitions depends on the concrete number of processes in
each particular instantiation.

The main contribution of this paper is the principle or parametrized
invariance, presented as proof rules that capture the effect of
single steps of threads involved in the property and extra
\emph{arbitrary} threads.
%
%
The parametrized invariance rules automatically discharge a
\emph{finite} number of VCs, whose validity imply the correctness for
all system instantiations.  For simplicity we present the rules for
fully symmetric systems (in which thread identifiers are only compared
with equality) and show that all VCs generated are
\emph{quantifier-free} (as long as transition relations and
specifications are quantifier-free, which is the case is conventional
system descriptions).

For many data-types one can use directly SMT
solvers~\cite{moura08Z3,ganzinger04dpllt}, or specialized decision
procedures built on top.  We show here how to use the decision
procedure for a quantifier-free theory of single linked list layouts
with locks~\cite{sanchez10decision} to verify fine-grained concurrent
list implementation.  Other powerful logics and tools for building
similar decision procedures
include~\cite{lahiri08back,madhusudan11decidable}.


\noindent\myparagraph{Related Work}
The problem of uniform verification of parametrized systems has
received a lot of attention in recent years.  This problem is, in
general, undecidable~\cite{apt86limits}, even for finite state
components. There are two general ways to overcome this limitation:
deductive proof methods as the one we propose here, and (necessarily
incomplete) algorithmic approaches.

Most algorithmic methods are restricted to finite state
processes~\cite{clarke86reasoning, clarke87avoiding,emerson00reducing}
to obtain decidability. Examples are synchronous communication
processes~\cite{german92reasoning,emerson96automatic}; systems with
only conjunctive guards or only disjunctive
guards~\cite{emerson00reducing}; implicit
induction~\cite{emerson95reasoning}; network
invariants~\cite{lesens97automatic}; etc.  A related technique, used
in parametrized model checking, is symmetry
reduction~\cite{emerson96symmetry,clarke96exploiting}. 
A very powerful method is invisible
invariants~\cite{pnueli01automatic,zuck04model,arons01parameterized},
which works by generating invariants on small instantiations and
generalizing these to parametrized invariants. However, this method is
so far also restricted to finite state processes.

A different tradition of automatic (incomplete) approaches is based on
abstracting control and data altogether, for example representing
configurations as words from a regular
language~\cite{abdulla99handling,bozzano02beyond,abdulla08approximated,kesten98algorithmic}
Other approaches use abstraction, like thread
quantification~\cite{berdine08thread} and environment
abstraction~\cite{clarke08environment}, based on similar principles as
the full symmetry presented here, but relying on building specific
abstract domains that abstract symbolic states instead of using SMT
solvers.

In contrast with these methods, the verification framework we present
here can handle infinite data. The price to pay is, of course,
automation because one needs to provide some support invariants.  We
see our line of research as complementary to the lines mentioned
above. We start from a general method and investigate how to improve
automation as opposed to start from a restricted automatic technique
and improve its applicability.  The VCs we generate can still be
verified automatically as long as there are decision procedures for
the data that the program manipulates.

Our target application is the verification of concurrent
datatypes~\cite{herlihy08art}, where the main difficulty arises from
the mix of \emph{unstructured} \emph{unbounded} concurrency and heap
manipulation. Unstructured refers to programs that are not structured
in sections protected by locks but that allow a more liberal pattern
of shared memory accesses. Unbounded refers to the lack of bound on
the number of running threads. Concurrent datatypes can be modeled
naturally as fully symmetric parametrized systems, where each thread
executes in parallel a client of the datatype.
Temporal deductive methods~\cite{manna95temporal}, like ours, are very
powerful to reason about (structured or unstructured) concurrency.



The rest of the paper is structured as follows.
Section~\ref{sec:prelim} includes the preliminaries.
Section~\ref{sec:rules} introduces the parametrized invariance rule.
Section~\ref{sec:example} contains the examples, a description of
our tool and empirical evaluation results. Finally, Section~\ref{sec:conclusion} 
concludes.

\section{Running Example and Preliminaries}
\label{sec:prelim}

\begin{figure}[t]
	\hspace{-2em}\includegraphics[scale=0.35]{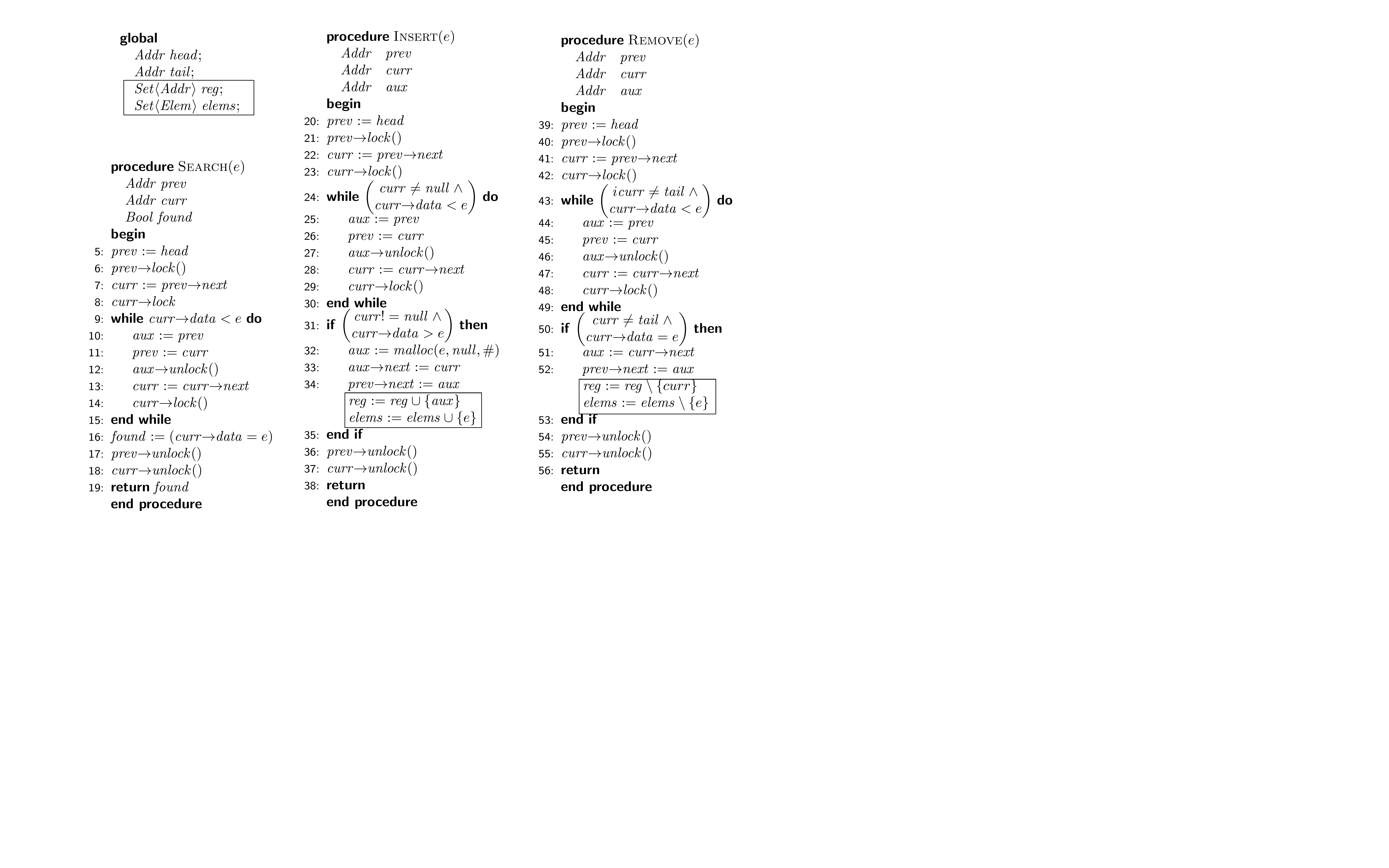}
	\caption{Lock-coupling single linked list implementation}
	\label{fig:listcode}
\end{figure}

\myparagraph{Running Example}
We will use as a running example a concurrent data-type that
implements a set~\cite{herlihy08art} using fine-grain locks, shown in
Fig.~\ref{fig:listcode}. Appendix~\ref{sec:appExamples} contains
simpler and more detailed examples of infinite state mutual exclusion
protocols.  Lock-coupling concurrent lists implement sets by
maintaining an ordered list of non-repeating elements. Each node in
the list stores an element, a pointer to the next node in the list and
a lock used to protect concurrent accesses. To search an element, a
thread advances through the list acquiring a lock before visiting a
node. This lock is only released after the lock of the next node has
been acquired.  Concurrent lists also maintain two sentinel nodes,
\head and \tail, with phantom values representing the lowest and
highest possible values, $-\infty$ and $+\infty$ respectively. Sentinel
nodes are not modified at runtime.
We define two ``ghost'' variables that aid the verification: \region,
a set of addresses that contains the set of address pointing to nodes
in the list; 
\noindent\begin{wrapfigure}[12]{l}{15em}
  \centering
  \vspace{-1.5em}
   \begin{tabular}{c}
	   \includegraphics[scale=0.35]{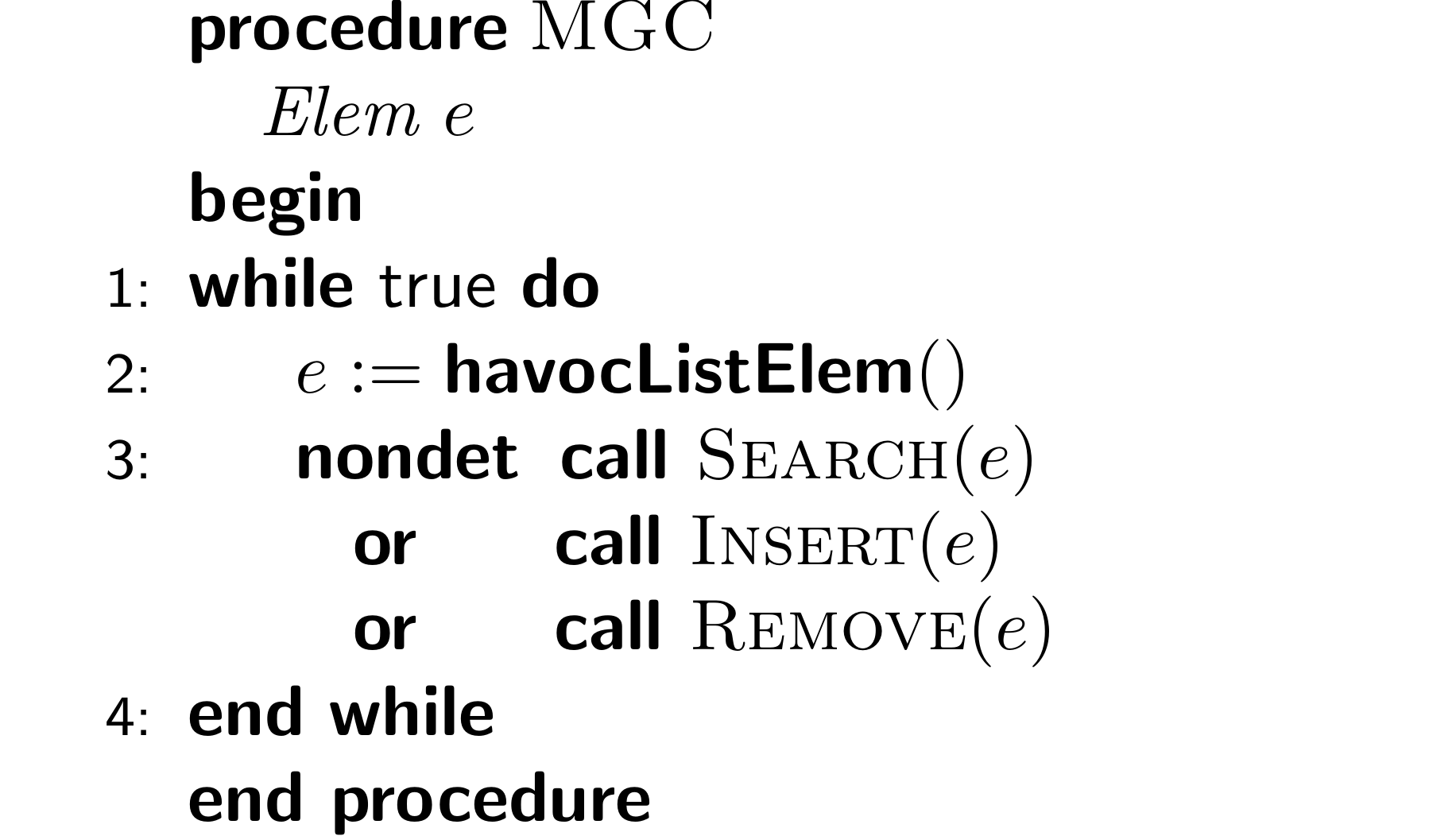}
   \end{tabular}
  \caption{Most General Client}
  \label{fig:mgc}
\end{wrapfigure}
and \elements, a set of elements we use to keep track of
elements contained in the list. 
Ghost variables are compiled away and
are only used in the verification process. In Fig.~\ref{fig:listcode}
ghost variables and code appear inside a box.
As lock-coupling lists implement sets, three main operations are provided:
\begin{inparaenum}[(a)]
\item \Search: finds an element in the list;
\item \Insert: adds a new element to the list; and
\item \Remove: deletes an element in the list.
\end{inparaenum}
For verification purposes, it is common to define the most general
client \MostGeneralClient shown in Fig.~\ref{fig:mgc}.
Each process in the parametrized system runs \MostGeneralClient
choosing non-deterministically a method and its parameters.

\begin{figure}[t]
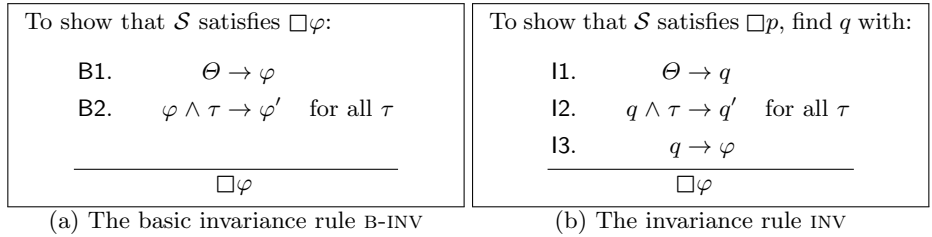

  \begin{tabular}{cc}
    \bINVRule & \INVRule \\
    (a) The basic invariance rule \bINV & (b) The invariance rule \INV
  \end{tabular}
  \caption{Rules \bINV and \INV for non-parametrized systems.}
  \label{fig:inv}
\end{figure}

\myparagraph{Preliminaries} Our verification task starts from a
program, and a safety property described as a state predicate. A
system is correct if all states in all the traces of the transition
system that models the set of executions of the program satisfy the
safety property.
%
%
%

A transition system is a tuple $\Sys:\tupleof{\V, \Theta, \T}$ where
\V is a finite set of (typed) variables, $\Theta$ is a first-order
assertion over the variables which describes the possible initial
states, and \T is a finite set of transitions.  We model program data
using multi-sorted first order logic. A signature $\Sigma:(S,F,P)$
consists of a set of sorts $S$ (corresponding to the types of the data
that the program manipulates), a set $F$ of function symbols, and a
set $P$ of predicate symbols.  We use $\sigProg$ for the signature of
the datatypes in a given program and $\TProg$ for the theory that
allows to reason about formulas in $\sigProg$.  A \emph{state} is an
interpretation of \V that assigns a value of the corresponding type to
each program variable.  A transition is represented by a logical
relation $\tau(s,s')$ that describes the relation between the values
of the variables in a state $s$ and a successor state $s'$.  A
\emph{run} of $\Sys$ is an infinite sequence
$s_0\tau_0s_1\tau_1s_2\ldots$ of states and transitions such that
\begin{inparaenum}[(a)]
\item the first state is initial: $s_0\vDash \Theta$;
\item all steps are legal: $\tau_i(s_i,s_{i+1})$, that is, $\tau_i$ is
  taken at $s_i$, leading to state $s_{i+1}$.
\end{inparaenum}
%

A system $\Sys$ satisfies a safety property $\Always \varphi$, which
we write $\Sys\vDash\Always \varphi$, whenever all runs of $\Sys$
satisfy $\varphi$ at all states.
For non-parametrized systems, invariants can be proved using the
classical invariance rules~\cite{manna95temporal}, shown in
Fig.~\ref{fig:inv}. The basic rule \bINV establishes that if the
candidate invariant $\varphi$ holds initially and is preserved by
every transition then $\varphi$ is indeed an invariant. Rule \INV uses
an intermediate strengthening invariant $q$. If $q$ implies $\varphi$
and $q$ is an invariant, then $\varphi$ is also an invariant. For
non-parametrized systems, the premises in these rules discharge a
number of verification conditions linear in the number of transitions.
To use these invariance rules for parametrized systems, one either
needs to use quantification or discharge an unbounded number of VCs,
depending on the number of processes.

\subsubsection{Parametrized Concurrent Programs.}

Parametrized programs consist of the parallel execution of process
running the same program (the extension to an unbounded number of
processes each running a program from a finite collection is
trivial). We assume asynchronous interleaving semantics for parallel
composition.  A program is described as a sequence of statements, each
assigned to a program location in the range $\Loc:1\ldots{}L$. Each
instruction can manipulate a collection of typed variables partitioned
into $\Vglobal$, the set of \emph{global} variables, and $\Vlocal$,
the set of \emph{local} variables.
%
%
There is one special local variable $\pc$ of sort \Loc that stores the
program counter of each thread.  For example, for the program in
Fig.~\ref{fig:listcode}, \TProg is the combination of \TLLThree (the
theory of single-linked lists in the heap with
locks~\cite{sanchez10decision}), combined with finite discrete values
(for program locations).
%
%
%
%
%
%
%
%
%
In transition relations we use a primed variable $v'$ to denote the
value of variable $v$ after a transition is taken.

A parametrized program $P$ is associated with a parametrized system
$\Sys$, a collection of transition systems $\Sys[M]$, one for each
number of running threads. We use $[M]$ to denote the set
$\{0,\ldots,M-1\}$ of concrete thread identifiers. For each $M$, there
is a system $\Sys[M]:\tupleof{\V,\Theta,\T}$ consisting of:
\begin{compactitem}
  \item 
    The set \V of variables is $\Vglobal\cup\{v[k]\}\cup\{\pc[k]\}$
    where there is one $v[k]$ for each $v\in\Vlocal$ and for each
    $k\in[M]$, and one $\pc[k]$ for each $k\in[M]$.
%
%
   
  \item An initial condition $\Theta$, which is described by two
    predicates $\ThetaGlobal$ (that only refers to variables from
    \Vglobal) and $\ThetaLocal$ (that can refer to variables in
    \Vglobal and \Vlocal).
%
%
    Given a thread identifier $a\in[M]$ for a concrete system
    $\Sys[M]$, $\ThetaLocal[a]$ is the initial condition for thread
    $a$, obtained by replacing $v[a]$ for every occurrence of $v$ in
    $\ThetaLocal$.
  \item $\T$ contains a transition $\tau_{\ell}[a]$ for each location
    and thread $a$ in $[M]$ obtained from $\tau_{\ell}$ by replacing
    every occurrence of $v$ by $v[a]$, and of $v'$ by $v'[a]$.
\end{compactitem}

We use $\V^t$ to denote all variables of sort $t$ in set \V.

 \begin{example}
   Consider the lock-coupling list program in Fig.~\ref{fig:listcode}.
	 The instance of this program consisting of two running threads
   contains the following variables:
%
\begin{compactitem}
\item[]
$
   \begin{array}{llllll}
     \V = \{ \head, \tail, \region, \elements, & e[0], & \prev[0], & \curr[0], & \aux[0], & \found[0], \\
     & e[1], &\ prev[1], & \curr[1], & \aux[1], & \found[1] \} 
     \end{array} 
$
\end{compactitem}
There are $118$ transitions in $\MostGeneralClient[2]$, $59$
transitions for each thread, one for each line in the program. For
non-parametrized systems, like $\MostGeneralClient[2]$, we use the
predicate $\pres$ in transition relations to list the variables that
are not changed by the transition. That is $\pres(\head,\tail)$ is
simply a short for $\head'=\head\;\And\;\tail'=\tail$.

 %
 %
   \label{ex:listsInstance}
 \end{example}

\section{Parametrized Formulas and Parametrized Proof Rules}
\label{sec:rules}

We show in this paper how to specify and prove invariant properties of
parametrized systems. Unlike in~\cite{pnueli01automatic} we generate
quantifier-free verification conditions, enabling the development of
decision procedures for complex datatypes.

To model thread ids we introduce the sort $\sTID$ interpreted as an
unbounded discrete set. The signature $\sigTID$ contains only $=$ and
$\neq$, and no constructor.  We enrich $\TProg$ using the theory of
arrays $\TArray$~(see \cite{bradley06what}) with indices from $\sTID$
and elements ranging over sorts $t$ from the local variables of
$\TProg$. For each local variable $v$ of type $t$ in the program, we
add a global variable $a_v$ of sort $\sArray\tupleof{t}$, including
$a_\pc$ for the program counter $\pc$.  The expression $a_v(k)$
denotes the element of sort $t$ stored in array $a_v$ at position
given by expression $k$ of sort $\sTID$. The expression
$a_v\{k\leftarrow e\}$ corresponds to an array update, and denotes the
array that results from $a_v$ by replacing the element at position $k$
with $e$. To simplify notation, we use $v(k)$ for $a_v(k)$, and
$v\{k\leftarrow e\}$ for $a_v\{k \leftarrow e\}$. Note how $v[0]$ is
different from $v(k)$: the term $v[0]$ is an atomic term in $V$ (for a
concrete system $\Sys[M]$) referring to the local program variable $v$
of a concrete thread with id $0$. On the other hand, $v(k)$ is a
non-atomic term built using the signature of arrays, where $k$ is a
variable (logical variable, not program variable)  of sort $\sTID$.


\newcommand{\Vparam}{\ensuremath{V_{\textit{param}}}\xspace}
\newcommand{\Vbaseparam}{\ensuremath{V_{\textit{bsp}}}\xspace}

Variables of sort $\sTID$ indexing arrays play a special role, so we
classify formulas depending on the sets of variables used. The
parametrized set of program variables with index variables $X$ of sort
$\sTID$ is:
\[
\Vparam(X) = \Vglobal \cup \{ a_v\;|\; v\in \Vlocal \} \cup \{ a_\pc \} \cup X
\]
We use $T$ for the union of theories $\TProg$, $\TTID$ and $\TArray$.
$F_T(X)$ is the set of first-order formulas constructed using
predicates and symbols from $T$ and variables from $\Vparam(X)$.
Given a \sTID variable $k$ and a program statement, we construct the
parametrized transition relation as before, but using array reads and
updates (to position $k$) instead of concrete local variable reads and
updates. For parametrized formulas, the predicate $\pres$ is defined
with array extensional equality for unmodified local variables.



We similarly define the parametrized initial condition for a given set
of thread identifiers $X$ as:
\[ \Theta(X) \;:\; \ThetaGlobal \And \bigwedge\limits_{k\in X} \ThetaLocal(k) \]
where $\ThetaLocal(k)$ is obtained by replacing every local variable
$v$ in $\ThetaLocal$ by $v(k)$. 



A \emph{parametrized formula} $\TFof{\overline{k}}$ with free
variables $\overline{k}=(k_0,\ldots, k_n)$ of sort \sTID is a formula
from $F_T(\{k_0,\ldots,k_n\})$.  Note, in particular, how parametrized
formulas cannot refer to any constant thread identifier.  We use
$\Var(\varphi)$ for the set of free $\sTID$ variables in $\varphi$.

Given a concrete number of threads $N$, a \emph{concretization} of
expression $p(\overline{k})$ is characterized by a substitution
$\alpha:\overline{k}\Into[N]$ that assigns to each variable in
$\overline{k}$ a unique constant thread identifier in the instance
system $\Sys[N]$. The application of $\alpha$ for expressions $p$ is
defined inductively, where the base cases are:
\[ \begin{array}{r@{\;\;}c@{\;\;}l}
  \alpha(v(k_i)) &\mapsto& v[\alpha(k_i)] \\
  \alpha(w=v\{k_i\leftarrow e\}) &\mapsto& \big(w[\alpha(k_i)]=e \And \bigwedge\limits_{a\in N\setminus \alpha(k_i)} w[a]=v[a]\big) 
\end{array}
\]
Essentially, a concretization provides the state predicate for system
$\Sys[N]$ that results from $p(\overline{k})$ by instantiating
$\overline{k}$ according to $\alpha$. 

We can formulate the uniform verification problem in terms of
concretizations. Given a parametrized system $\Sys$, a universal
safety property of the form $\forall\overline{k} \;.\;\Always
p(\overline{k})$ holds whenever for every $N$ and substitution
$\alpha:\overline{k}\Into [N]$, the concrete closed system $\Sys[N]$
satisfies $\Sys[N]\vDash\Always\alpha(p(\overline{k}))$.  In this case
we simply write $\Sys\vDash \Always p$ and say that $p$ is a
parametrized invariant of $\Sys$.

%

A na\"{i}ve approach to prove parametrized inductive invariants is to
enumerate all instances and repeatedly use rule \INV for each
one. However, this approach requires proving an unbounded number of
verification conditions because one (potentially different) VC is
discharged per transition and thread in every instantiated closed
system.
\begin{figure}[t!]
  \centering
  \pINVRule
  \caption{The parametrized invariance rule \pINV}
  \label{fig:pinv}
\end{figure}

\subsubsection{Parametrized Proof Rules.}
We introduce here specialized proof rules for parametrized systems,
which allow to prove parametrized invariants discharging only a finite
number of verification conditions.  Rule $\pINV$ in
Fig.~\ref{fig:pinv} presents the basic parametrized invariance
rule. Premise \Premise{P1} guarantees that the initial condition holds
for all instantiations.  Premise \Premise{P2} guarantees that
$\varphi$ is preserved under transitions of the threads referred in
the formula, and \Premise{P3} guarantees that $\varphi$ is preserved
under transitions \emph{of any other thread}.  \Premise{P1} discharges
only one verification condition, \Premise{P2} discharges one VC per
transition in the system and per index variable in the formula
$\varphi$. Finally, \Premise{P3} generates one extra VC per transition
in the system. All these VCs are quantifier-free provided that
$\varphi$ is quantifier-free.  The following theorem justifies the
introduction of rule \pINV:


\newcounter{thm-soundness}
\setcounter{thm-soundness}{\value{theorem}}

\begin{theorem}[Soundness]
  \label{thm:soundness}
  Let $\Sys$ be a parametrized system and $\Always \varphi$ a parametrized
  safety property. If \Premise{P1}, \Premise{P2} and \Premise{P3}
  hold, then $\Sys\vDash\Always \varphi$.
\end{theorem}

\begin{proof} (sketch) The proof proceeds by contradiction, assuming
  that the premises hold but $\Sys\not\vDash\Always \varphi$.  There must be
  an $N$ and a concretization $\alpha$ for which
  $\Sys[N]\not\vDash\Always \alpha(\varphi)$. Hence, by soundness of the
  \INV{} rule for closed systems, there must be a premise of \INV{}
  that is not valid. By cases, one uses the counter-model of the
	offending premise to build a counter-model of the corresponding
  premise in \pINV.
  \qed
\end{proof}

There are cases in which premise~\Premise{P3} cannot be proved, even
if $\varphi$ is initial and preserved by all transitions of all
threads.
The reason is that, in the antecedent of \Premise{P3}, $\varphi$ does
not refer to the fresh arbitrary thread introduced.  In other words,
\pINV tries to prove the property for an arbitrary process at all
reachable system states without assuming anything about any other
thread. It is sound, however, to assume in the pre-state and \emph{for
  all processes} the property one intends to prove. The notion of
\emph{support} allows to strengthen the antecedent to refer to all
threads involved in the verification condition, including the fresh
new thread.

\begin{definition}[support]
  Let $\psi$ be a parametrized formula (the support) and let $(A\Into
  B)$ be a parametrized formula with $\Var(A\Into B)=X$. We say
  that $\psi$ supports $(A\Into B)$, whenever 
  \( \big[ \big(\bigwedge_{\sigma\in S} \psi\sigma \And A\big) \Into B \big] \)
  is valid,
  where $S$ is a subset of the partial substitutions
  $\Var(\psi)\partialmap X$.  
\end{definition}

We use $\psi\supports(A\Into B)$ as a short for
$(\big(\bigwedge_{\sigma\in S} \psi\sigma \And A\big) \Into B)$.
%
%
%
%
%
We can strengthen premise \Premise{P3} with self-support, so $\varphi$
can be assumed (in the pre-state) for every thread, in particular for
the fresh thread that takes the transition:

\vspace{0.8em}
\pINVThree
\vspace{0.8em}

For example, let $\varphi(i)$ be a candidate invariant with one thread
variable (an index 1 invariant candidate). Premise~\Premise{P3'} is \(
\big(\varphi \supports (j\neq i \And \tau^{(j)} \Into
\varphi'(i))\big) \), or equivalently%
\[ \big(\varphi(j) \And \varphi(i) \And j\neq i \And \tau^{(j)}\big) \Into \varphi'(i). \]
Note how $\varphi(j)$ in the antecedent is the result of instantiating
$\varphi$ for the fresh thread $j$.
Rule \pINV can fail to prove invariants if they are not inductive. As
for closed systems, one needs to strengthen invariants. However, it is
not necessary the case that by conjoining the candidate and its
strengthening one obtains a \pINV inductive invariant. Instead, one
needs to use a previously proved invariant as \emph{support} to
consider also freshly introduced process identifiers. This idea is
captured by rule \spINV in Fig.~\ref{fig:spINV}.
\begin{theorem}
  \label{thm:soundnessStr}
  Let $\Sys$ be a parametrized system and $\Always \varphi$ a
  parametrized safety property. If \Premise{S0}, \Premise{S1},
  \Premise{S2} and \Premise{S3} hold, then $\Sys\vDash\Always \varphi$.
\end{theorem}

\begin{figure}[t]
  \strgINVRule 
  \caption{The general strengthening parametrized invariance rule
    \spINV.}
  \label{fig:spINV}
\end{figure}

\subsubsection{Graph Proof Rules}

We now introduce a final specialized proof rule for parametrized
systems. When using \spINV, \Premise{S0} requires to start from an
already proved invariant. However, in some cases invariants mutually
depend on each other.  For example, in the proof of shape preservation
of concurrent single-linked list programs, like the one in
Fig.~\ref{fig:listcode}, one requires that the pointers \curr and
\prev used in the list traversal do not alias. This fact depends on
the list having at all program states the shape of a non-cyclic list.
\begin{figure}[t]
	\GeneralGraphINVRule \caption{The graph parametrized invariance rule \gINV.}
	\label{fig:gINV}
\end{figure}
A correct but na\"{i}ve solution would be to write down all necessary
conditions as a single formula and prove it invariant using
\pINV. Unfortunately, this approach does not scale when using
sophisticated decision procedures for infinite memory.  A more
efficient approach consists on building the proof modularly, splitting
the invariant into meaningful subformulas to be used when
required. Modularity motivates the introduction of \gINV, a rule for
proof graphs shown in Fig.~\ref{fig:gINV}. This rule handles cases in
which invariants that mutually dependent on each other need to be
verified.

A proof graph $(V,E)$ has candidate invariants as nodes. An edge
between two nodes indicates that in order to prove the formula pointed
by the edge it is useful to use the formula at the origin of the edge
as support. As a particular case, a formula with no incident edges is
inductive and can be shown with \pINV.

\begin{theorem}
  \label{thm:soundnessGraph}
  Let $\Sys$ be a parametrized system and $(V,E)$ a proof graph. If
  \Premise{G1}, \Premise{G2}, and \Premise{G3} hold, then
  $\Sys\vDash\Always \psi$ for all $\psi\in V$.
\end{theorem}

\begin{proof}
  By contradiction assume that some formula in $V$ is not an
  invariant. Then, consider a shortest path to a violation in any
  concrete system $\Sys[M]$. Let $\psi\in V$ be the violated
  formula. By $\Premise{G_1}$, the path cannot be empty because
  $\Premise{G_1}$ implies initiation of all formulas in $V$ for all
  concrete system instances.  Hence, the offending state $s$ violating
  $\psi$ has a predecessor state $s_{\text{pre}}$ in the path, which
  by assumption, satisfies all formulas in $V$, and in particular all
  formulas in $\{ \psi_i\;|\;\psi\in E\}$ \ie with outgoing edges
  incident to $\psi$. Premises $\Premise{G2}$ and $\Premise{G3}$,
  guarantee that the execution step from $s_{\text{pre}}$ to $s$
  guarantees $\psi$ in $s$, which is a contradiction. \qed
\end{proof}

We now show not that for fully symmetric systems, the dependencies
with arrays in the parametrized formulas can be eliminated preserving
validity, generating formulas that decision procedures can reason
about.
\begin{theorem}[Concretization]
  Let $\varphi(\overline{k})$ be with $|\overline{k}|=n$. Then
  $\varphi(\overline{k})$ is valid if and only if
  $\bigwedge\limits_{\alpha\in A} \alpha(\varphi)$ is valid where $A$
  is the set of all possible assignments of variables in
  $\Var(\varphi)$ to $[n]$.
\end{theorem}

For example, if one intends to prove that $p(i)$ is inductive, the
concretization theorem allows to reduce \Premise{P3} in \pINV to
$(p[0] \And \tau[1] \Into p'[0])$, where $p[0]$ is a short for
$\alpha(p(i))$ with $\alpha:i\Into 0$. This formula involves no
arrays. Similarly, to show $\Always p(i)$ with support invariant
$q(j)$, rule \Premise{S3} can be reduced to:
\[
\begin{array}{rcl}
   q[0] \And q[1]  \And p[0] \And p[1] \And \tau[1] &\Into& p'[0]\\
\end{array}
\]

In practice, the concretization can be performed upfront before
discharging the verification condition to the SMT-Solver, or handled
using the theory of uninterpreted functions and let the solver perform
the search and propagation.


\section{Implementation and Empirical Evaluation}
\label{sec:example}

We illustrate the use of our parametrized invariance rules proving
list shape preservation and some functional properties about set
representation of the concurrent list implementation presented in
Fig.~\ref{fig:listcode}. We also show mutual exclusion for some
infinite state protocols that use integers and sets of integers (see
the appendix~\label{sec:appExamples} for details).

The proof rules are implemented in the temporal theorem prover tool
\Leap, under development at the IMDEA Software
Institute\footnote{Available at \url{http://software.imdea.org/leap}}.
\Leap parses a temporal specification and a program descriptions in a
C-like language.  \Leap automatically generates VCs applying the
parametrized invariance rules presented in this paper. The validity of
each VC is then verified using a suitable decision procedure (DP) for
each theory.

We compare here three decision procedures built on top the SMT solvers
Z3 and Yices:
\begin{inparaenum}[(1)]
\item a simple DP that can reason only about program locations, and
  considers all other predicates as uninterpreted;
\item a DP based on \TLLThree capable of reasoning about single-linked
  lists layouts in the heap with locks to aid in the verification of
  fine-grain locking algorithms; and
\item a DP that reasons about program locations, integers and finite
  sets of integers with minimum and maximum functions (for the mutual
  exclusion protocols).
\end{inparaenum}
The last two decision procedures and their implementation are based on
small model theorems. The satisfiability of a quantifier free formula
is reduced to the search for a model (up to a sufficiently large
size).
\Leap also implements some heuristic optimizations (called
\emph{tactics}) like attempting first to use a simpler decision
procedure or instantiating support lazily. This speeds the solvers in
many valid instances by reducing the formulas obtained by partial
assignments in the application of rules \spINV or \gINV.  
%

\myparagraph{List Preservation and Set Representation for Concurrent
  Lists} 

We prove that the program in Fig.~\ref{fig:listcode} satisfies:
\begin{inparaenum}[(1)]
\item list shape preservation; and
\item the list implements a set, whose elements correspond to those
  stored in \elements.
\end{inparaenum}
The theory \TLLThree (see~\cite{sanchez10decision}) allows to reason
about addresses, elements, locks, sets, order, cells (\ie list nodes),
memory and list reachability.  A cell is a struct containing an
element, a pointer to next node in the list and lock to protect the
cell. A lock is associated with operations $\lock$ and $\unlock$ to
acquire and release. The memory (\heap) is modeled as an array of
cells indexed by addresses.
The specification is:
\[
	\formulaFullListReduced
\]
Formula \phiPreserve is 0-index since it only constrains global
variables. (L1) establishes that \fNull belongs to \region and that
\region is exactly the set of addresses reachable in the \heap
starting from \head, which ensures that the list is acyclic. (L2) and
(L3) express some sanity properties of the sentinel nodes \head and
\tail.  Finally, (L4) establishes that \elements is the set of
elements in cells referenced by addresses in \region, and that the
list is ordered. The main specification is \invPreserve, defined as
$\Always \phiPreserve$.

Using \pINV, \Leap can establish that \invPreserve holds initially,
but fails to prove that \invPreserve is preserved by all transitions.
The use of decision procedures for proving VCs allows to obtain
counter-examples as models of an execution step that leads to a
violation of the desired invariant. \Leap parses the counterexample
(model) returned by the SMT solver, which is usually very small,
involves only few threads and allows to understand the missing
intermediate facts.  In practice, these models allow to write easily
the support invariants.  We introduce some support invariants that
allow to prove \invPreserve.

Invariant $\invRegion(i)$ describes that local variables $\prev$,
$\curr$ and $\aux$ point to cells within the region of the list
$\region$, and that these variables cannot be null or point to $\head$
or $\tail$.  The formula $\invRegion$ is $1$-index (because it needs to
refer to local variables of a single thread).
Invariant $\invNext(i)$ captures the relative position in the list of
the cells pointed by \head and \tail and local variables \prev, \curr
and \aux. This invariant is needed for (L2).
To prove (L3) and (L4) we need to show that order is preserved. We
introduce $\invOrder(i)$, which captures the increasing order between
the data in cells pointed by $\curr$, $\prev$ and $\aux$ and by the
searched, inserted or removed element $e$.
Invariant $\invLock(i)$ captures those program locations at which a thread
owns a cell in the heap by an acquired lock.
Finally, $\invDisjoint (i,j)$, defined as $\Always \phiDisjoint(i,j)$
encodes that the calls to \emph{malloc} by different threads return
different addresses:
\[
	\formulaFullDisjoint
\]
%

Other properties verified for the concurrent list are functional like
specifications.
Invariant $\invFunSchLinear(i)$ establishes that the result of \Search matches 
with the presence of the searched element $e$ at \Search's linearization point;
$\invFunSchInsert(i)$ states that if a search is successful then $e$ was 
inserted earlier in the history; and
$\invFunSchRemove(i)$ captures the fact that if the search is unsuccessful then 
either $e$ was never inserted or it was removed, and it was not present at the 
linearization point of \Search.
The invariants $\invFuncRemove(i)$, $\invFuncInsert(i)$ and $\invFuncSearch(i)$ 
consider the case in which one thread handles different elements than all other 
threads. In this case, the specification is similar to a sequential functional
specification (an element is found if and only if it is in the list, an element 
is not present after removal and an element is present after insertion).

\myparagraph{Infinite State Mutual Exclusion Protocols} 

We also report the proof of mutual exclusion of some simple infinite
state protocols that use tickets. The first protocol uses two global
integer variables, one to store the next available ticket, and another
to represent the minimum ticket present. The decision procedure used
is Presburger arithmetic. The second protocol stores the tickets in a
global set of integers, and queries for the minimum element in the
set. The decision procedure used is Presburger Arithmetic combined
with finite sets of integers with minimum.

%
%

\begin{wrapfigure}[9]{l}{12.8em}
  \centering
  \vspace{-1.5em}
   \begin{tabular}{c}
	   \includegraphics[scale=0.4]{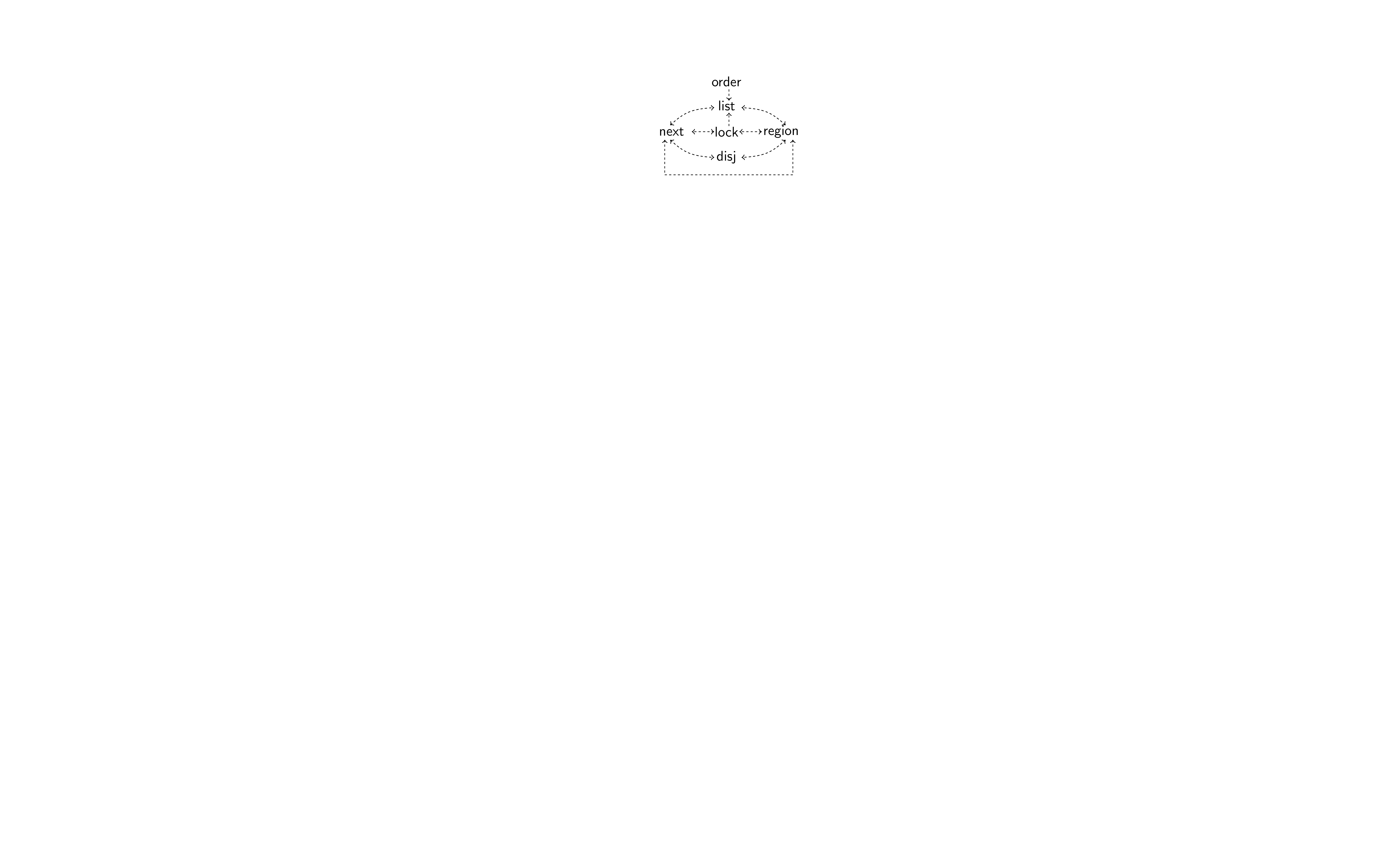}
   \end{tabular}
  \caption{Invariant dependencies}
  \label{fig:proof-graph}
\end{wrapfigure}

Fig.~\ref{fig:proof-graph} shows the proof graph encoding the proof of
\invPreserve.
%
\Leap can read proof graphs and apply \gINV.  Fig.~\ref{fig:results}
contains the results of this empirical evaluation, executed on a
computer with a 2.8 GHz processor and 8GB of memory.  Each row reports
the results for a single invariant. The first four columns show the
index of the formula, the total number of generated VCs, the number of
VCs proved by position, and the remaining VCs. The next four columns
show the total running time using the specialized decision procedures
with different tactics: ``Full supp'' corresponds to instantiating all
support invariants for all VCs; ``Supp'' corresponds to instantiate
only the necessary support;
\begin{figure}[t!]
	\centering
        \ResultsTable
  	\caption{VCs proved using each decision procedure and running times.}
        \vspace{-5em}
        \label{fig:results}
\end{figure}
``Offend'' corresponds to only using support in potentially offending
transitions; ``Tactics'' reports the
running time needed using some basic tactics like lazy 
instantiation and formula normalization and propagation.  TO
represents a timeout of 30 minutes. Our results
\noindent
indicates that, in practice, tactics
are essential for efficiency when handling non-trivial examples such
as concurrent lists.  Even though our decision procedures have room
for improvements, these results suggest that trying to compute an
over-approximation of the reachable state space for complicated
algorithms by iteratively computing formulas is not likely to be
feasible for complicated heap manipulating programs.


\section{Concluding Remarks}
\label{sec:conclusion}

This paper has introduced a temporal deductive technique for the
uniform verification problem of safety properties of infinite state
processes, in particular for the verification of concurrent datatypes
that manipulate data in the heap.
Our proof rules automatically discharge a finite collection of
verification conditions, which depend on the program description and
the diameter of the formula to prove, but not on the number of threads
in a particular instance. Each VC describes a small-step in the
execution of all corresponding instances.  The VCs are quantifier-free
as long as the formulas are quantifier free.  We use the theory of
arrays~\cite{bradley06what} to encode the local variables of a system
with an arbitrary number of threads, but the dependencies with arrays
can be eliminated, under the assumption of full symmetry. It is
immediate to extend our framework to a finite family of process
classes, for example to model client/server systems.

Future work includes invariant generation to simplify or even automate
proofs. We are studying how to extend the decision procedures with the
calculation of weakest precondition formulas
(like~\cite{lahiri08back}) and its use for parametrized systems
effectively to infer invariants, possibly from the target
invariant. We are also studying how to extend the ``invisible
invariant''
approach~\cite{pnueli01automatic,zuck04model,arons01parameterized} to
processes that manipulate infinite state, by instantiating small
systems with a few threads and limiting the exploration to only states
where data is limited in size as well. All candidate invariants
produced must then be verified with the proof rules presented here for
the general system.

We are also extending our previous work on abstract
interpretation-based invariant generation for parametrized
systems~\cite{sanchez12invariant} to handle complex datatypes. Our
work in~\cite{sanchez12invariant} was restricted to numerical domains.

Finally, another approach that we are currently investigating is to
use the proof rules presented here to enable a Horn-Clause
Verification engine~\cite{grebenshchikov12synthesizing} to
automatically generate parametrized invariants guided by the invariant
candidate goal. Our preliminary results are very promising but out of
the scope of this paper.

From a theoretical viewpoint the rule \spINV is complete (all
invariants can be proved by support inductive invariants), but the
proof of completeness is rather technical and is also out of the scope
of this paper.




\bibliographystyle{splncs03}



\bibliography{main,papers_list_short}

\begin{thebibliography}{10}
\providecommand{\url}[1]{\texttt{#1}}
\providecommand{\urlprefix}{URL }

\bibitem{abdulla99handling}
Abdulla, P.A., Bouajjani, A., Jonsson, B., Nilsson, M.: Handling global
  conditions in parametrized system verification. In: Proc. of CAV'99. pp.
  134--145 (1999)

\bibitem{abdulla08approximated}
Abdulla, P.A., Delzanno, G., Rezine, A.: Approximated parameterized
  verification of infinite-state processes with global conditions. FMSD  34(2),
   126--156 (2009)

\bibitem{apt86limits}
Apt, K.R., Kozen, D.C.: Limits for automatic verification of finite-state
  concurrent systems. Info. Proc. Letters  22(6),  307--309 (1986)

\bibitem{arons01parameterized}
Arons, T., Pnueli, A., Ruah, S., Xu, J., Zuck, L.D.: Parameterized verif. with
  automatically computed inductive assertions. In: Proc. of CAV'01. pp.
  221--234 (2001)

\bibitem{berdine08thread}
Berdine, J., Lev-Ami, T., Manevich, R., Ramalingam, G., Sagiv, S.: Thread
  quantification for concurrent shape analysis. In: Proc. of CAV'08. pp.
  399--413 (2008)

\bibitem{bradley06what}
Bradley, A.R., Manna, Z., Sipma., H.B.: What's decidable about arrays? In:
  VMCAI'06. LNCS, vol. 3855, pp. 427--442. Springer (2006)

\bibitem{clarke87avoiding}
Clarke, E.M., Grumberg, O.: Avoiding the state explosion problem in temporal
  logic model checking. In: PODC'87. pp. 294--303. ACM (1987)

\bibitem{clarke86reasoning}
Clarke, E.M., Grumberg, O., Browne, M.C.: Reasoning about networks with many
  identical finite-state processes. In: PODC'86. pp. 240--248. ACM (1986)

\bibitem{clarke96exploiting}
Clarke, E.M., Jha, S., Enders, R., Filkorn, T.: Exploiting symmetry in temporal
  logic model checking. FMSD  9(1/2),  77--104 (1996)

\bibitem{clarke08environment}
Clarke, E.M., Talupur, M., Veith, H.: Proving {Ptolemy} right: The environment
  abstraction framework for model checking concurrent systems. In: TACAS'08.
  LNCS, vol. 4963, pp. 33--47. Springer (2008)

\bibitem{emerson00reducing}
Emerson, E.A., Kahlon, V.: Reducing model checking of the many to the few. In:
  CADE'00. LNAI, vol. 1831, pp. 236--254. Springer (2000)

\bibitem{emerson95reasoning}
Emerson, E.A., Namjoshi, K.S.: Reasoning about rings. In: POPL'95. pp. 85--94.
  ACM (1995)

\bibitem{emerson96automatic}
Emerson, E.A., Namjoshi, K.S.: Automatic verification of parameterized
  synchronous systems. In: Proc. of CAV'96. LNCS, vol. 1102, pp. 87--98.
  Springer (1996)

\bibitem{emerson96symmetry}
Emerson, E.A., Sistla, A.P.: Symmetry and model checking. FMSD  9(1/2),
  105--131 (1996)

\bibitem{ganzinger04dpllt}
Ganzinger, H., Hagen, G., Nieuwenhuis, R., Oliveras, A., Tinelli, C.:
  {DPLL(T)}: Fast decision procedures. In: Proc. of CAV'04. pp. 175--188 (2004)

\bibitem{german92reasoning}
German, S.M., Sistla, A.P.: Reasoning about systems with many processes. J. of
  the ACM  39(3),  675--735 (1992)

\bibitem{grebenshchikov12synthesizing}
Grebenshchikov, S., Lopes, N.P., Popeea, C., Rybalchenko, A.: Synthesizing
  software veriﬁers from proof rules (2012)

\bibitem{herlihy08art}
Herlihy, M., Shavit, N.: The Art of Multiprocessor Programming.
  Morgran-Kaufmann (2008)

\bibitem{kesten98algorithmic}
Kesten, Y., Pnueli, A., on~Raviv, L.: Algorithmic verification of linear
  temporal logic specifications. In: ICALP'98. LNCS, vol. 1443, pp. 1--16.
  Springer (1998)

\bibitem{lahiri08back}
Lahiri, S.K., Qadeer, S.: Back to the future: revisiting precise program
  verification using {SMT} solvers. In: POPL'08. pp. 171--182. ACM (2008)

\bibitem{lesens97automatic}
Lesens, D., Halbwachs, N., Raymond, P.: Automatic verification of parameterized
  linear networks of processes. In: POPL'97. pp. 346--357. ACM (1997)

\bibitem{madhusudan11decidable}
Madhusudan, P., Parlato, G., Qiu, X.: Decidable logics combining heap
  structures and data. In: POPL'11. pp. 611--622. acm (2011)

\bibitem{manna95temporal}
Manna, Z., Pnueli, A.: Temporal Verif. of Reactive Systems. Springer (1995)

\bibitem{bozzano02beyond}
Marco~Bozzano, G.D.: Beyond parameterized verification. In: TACAS'02. LNCS,
  vol. 2280, pp. 221--235. Springer (2002)

\bibitem{moura08Z3}
de~Moura, L.M., Bj{\o}rner, N.: {Z3}: An efficient {SMT} solver. In: TACAS'08.
  LNCS, vol. 4963, pp. 337--340. Springer (2008)

\bibitem{pnueli01automatic}
Pnueli, A., Ruah, S., Zuck, L.D.: Automatic deductive verification with
  invisible invariants. In: TACAS'01. LNCS, vol. 2031, pp. 82--97. Springer
  (2001)

\bibitem{sanchez10decision}
S{\'a}nchez, A., S{\'a}nchez, C.: Decision procedures for the temporal
  verification of concurrent lists. In: ICFEM'10. LNCS, vol. 6447, pp. 74--89.
  Springer (2010)

\bibitem{sanchez12invariant}
S\'{a}nchez, A., Sankaranarayanan, S., S\'{a}nchez, C., Chang, B.Y.E.:
  Invariant generation for parametrized systems using self-reflection. In:
  SAS'12. LNCS, vol. 7460, pp. 146--163. Springer (2012)

\bibitem{zuck04model}
Zuck, L.D., Pnueli, A.: Model checking and abstraction to the aid of
  parameterized systems (a survey). Computer Languages, Systems \& Structures
  30,  139--169 (2004)

\end{thebibliography}

\vfill

\pagebreak 

\appendix

\section{Infinite State Mutual Exclusion Examples}
\label{sec:appExamples}

\myparagraph{Example: A Parametrized Mutual Exclusion Algorithm}
%
%

\begin{figure}[b!]
  \begin{tabular}{l@{\hspace{-1em}}l}
		\includegraphics[scale=0.35]{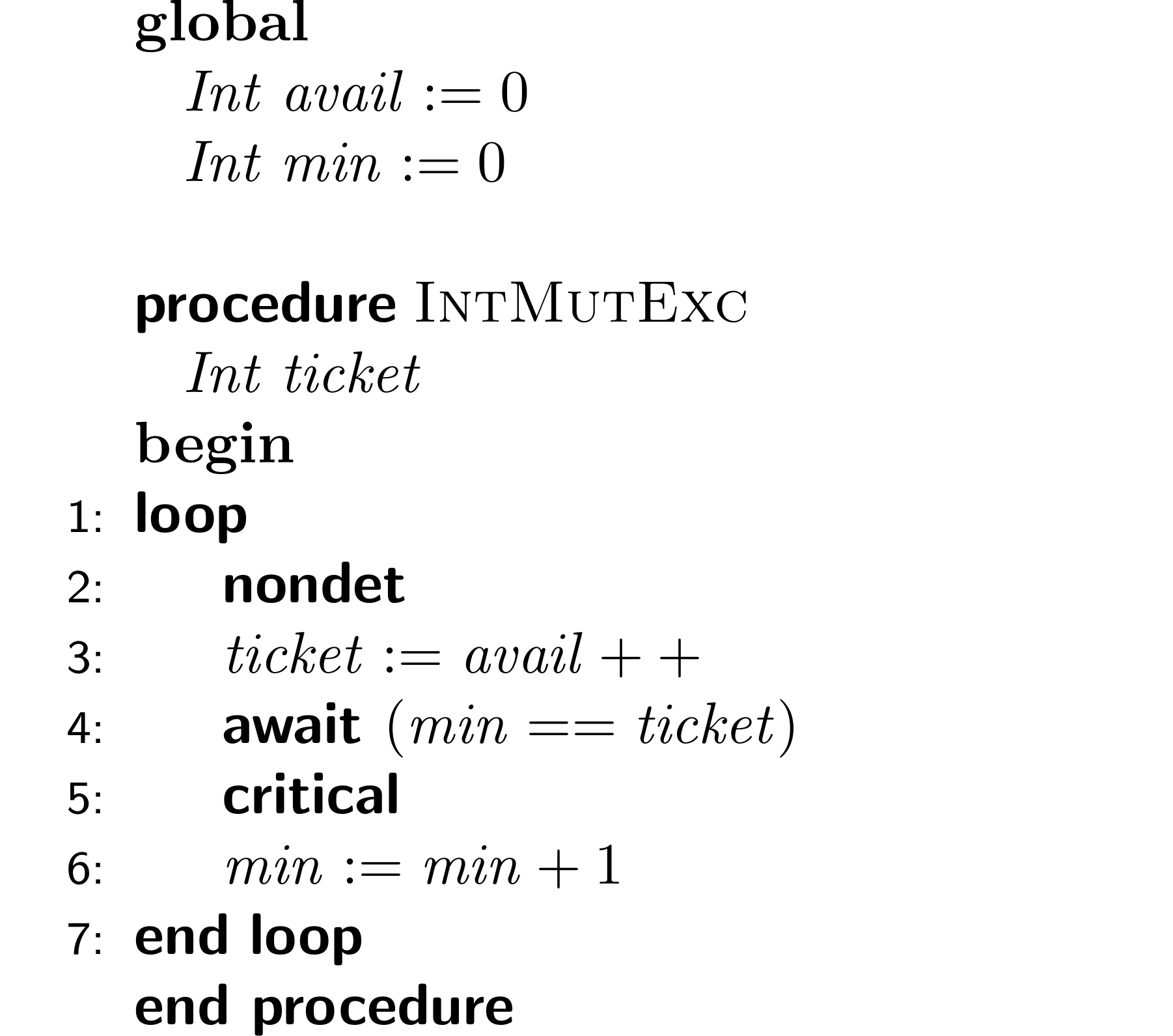} & 
		\includegraphics[scale=0.35]{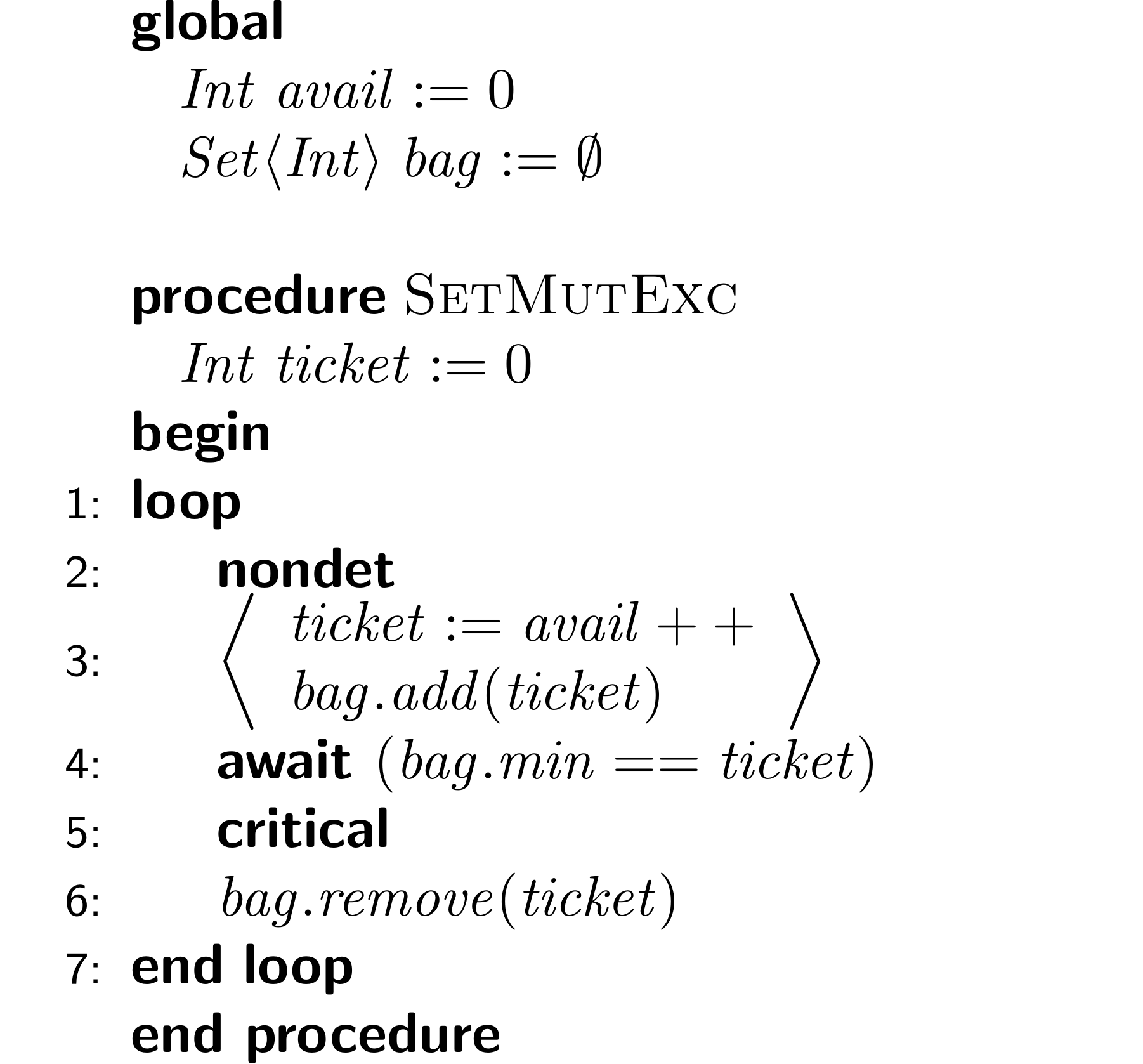} \\
    (a) \CriticalIntSectName, using two counters &
    (b) \CriticalSectName, using a set of integers
  \end{tabular}
  \caption{Two implementations of a ticket based mutual exclusion protocol}
  \label{fig:ticketmutex}
\end{figure}

Consider the program in Fig.~\ref{fig:ticketmutex}(b) which implements
mutual exclusion  using a simple
ticket-based protocol.  
Each thread that wants to access the critical section at line $5$,
acquires a unique increasing number (ticket) and announces its
intention to enter the critical section by adding the ticket to a
shared global set of tickets.  Then, the thread waits until its ticket
becomes the lowest value in the set before entering the critical
section. After a thread leaves the critical section it removes its
ticket from the set.  \CriticalSectName uses two global variables:
\avail, of type \Tint, which stores the shared counter; and \bag, of
type $\Tset\tupleof{\Tint}$, which stores the set of all threads that
are trying to access the critical section.  For any instance (number
of threads) the concrete system is an infinite state program, since
the available ticket is ever increasing. Program \CriticalIntSectName
in Fig.~\ref{fig:ticketmutex}(a) is similar except that is stores the
minimum value in a global variable of type \Tint.

\begin{example}
  Consider program \CriticalSectName in
  Fig.~\ref{fig:ticketmutex}(b). The instance consisting of two
  running threads, $\CriticalSectName[2]$, contains the following
  variables:
  \[ V = \{ \avail, \bag, \ticket[0], \ticket[1], \pc[0], \pc[1] \} \]
  Global variable \avail has type \Tint, and global variable \bag has
  type \TsetInt.  The instances of local variable $\ticket$ for
  threads $0$ and $1$, $\ticket[0]$ and $\ticket[1]$, have type \Tint.
  The program counters $\pc[0]$ and $\pc[1]$ have type $\Loc=\{1\ldots
  6\}$.  The initial condition of $\CriticalSectName[2]$ specifies
  that:
  \begin{equation}
  \begin{array}{l@{\hspace{3em}}rl}
    \Theta_g: \avail = 0 \;\land\; \bag = \emptyset &
    \Theta_l[0]: & \ticket[0] = 0\; \land\; \pc[0]=1 \\
    & \Theta_l[1]: & \ticket[1] = 0\; \land\; \pc[1]=1\\
  \end{array}
  \label{eq:iniconcrete}
  \end{equation}
  There are fourteen transitions in $\CriticalSectName[2]$, seven
  transitions for each thread: $\tau_1[0]\ldots\tau_7[0]$ and
  $\tau_1[1]\ldots\tau_7[1]$. The transitions corresponding to thread
  $0$ are:
 
  \[
  \begin{array}{r@{\;:\;}l@{$\;\land\;$}l@{$\;\land\;$}lcl}
    \tau_1[0]  & \pc[0]=1 & \pc'[0]=2 &  & & \pres(V \setminus \{\pc[0]\}) \\
    \tau_2[0]  & \pc[0]=2 & \pc'[0]=3 &  & & \pres(V \setminus \{\pc[0]\}) \\
%
    \tau_3[0]  & \pc[0]=3 & \pc'[0]=4 & \begin{pmatrix}\begin{array}{l}
      \ticket'[0]  = \avail \; \\
      \avail'   = avail+1    \\
      \bag'=\bag\cup\{\avail\} 
			\end{array}\end{pmatrix} & \land &
		\pres(\{\pc[1],\ticket[1]\}) \\
    \tau_4[0]  & \pc[0]=4 & \pc'[0]=5 & \bag.\setMin=\ticket[0]   &\;\land\;& \pres(V \setminus \{\pc[0]\}) \\
    \tau_5[0]  & \pc[0]=5 & \pc'[0]=6 & & &\pres(V \setminus \{\pc[0]\}) \\
    \tau_6[0]  & \pc[0]=6 & \pc'[0]=7 & \bag'=\bag\setminus\ticket[0] & \land &\pres(V \setminus \{\bag,\pc[0]\}) \\
    \tau_7[0]  & \pc[0]=7 & \pc'[0]=1 & & &\pres(V \setminus \{\pc[0]\})
  \end{array}
  \]
  The transitions for thread $1$ are analogous.  
%
%
  The predicate $\pres$ summarizes the preservation of variables'
  values. For example, in $\CriticalSectName[2]$, the predicate
  $\pres(V\setminus \{\bag,\pc[0]\})$ is simply:
  \[ 
  \avail'=\avail   \;\land\;    
  \ticket'[0]=\ticket[0] \;\land\; 
  \pc'[1]=\pc[1] \;\land\; \ticket'[1]=\ticket[1].
  \]
  \label{ex:mutextTwo}
\end{example}

\section{Empirical Evaluation: Mutual Exclusion}

\myparagraph{Mutual Exclusion for \normalfont{\CriticalIntSectName}:}

For the programs described in Fig.\ref{fig:ticketmutex} we use
$\Active(k)$ for $(\pc(k)=4,5,6)$ and $\Critical(k)$ for
$(\pc(k)=5,6)$. Mutual exclusion is specified as:
\[
\begin{array}{lcl}
	\mutex(i,j) & \defsym & \Always \big( i \neq j \Impl
    \lnot(\Critical(i) \land \Critical(j)) \big)
\end{array}
\]
Using the $\pINV$ rule to prove $\mutex$ fails for $\tau_4^{(i)}$, 
described as:
\[
	\mutex(i,j) \And
	\begin{pmatrix}
		\begin{array}{cc}
			\pc(i)=4 \And	\pc'=\pc\{i \leftarrow 5\} & \And \\
			\ticket(i) = \setMin & \And \\
			\pres(\tick, \setMin, \ticket(i), \ticket(j))
		\end{array}
	\end{pmatrix}
	\Impl \mutex'(i,j)
\]
The SMT Solver reports two counter models:
\[
	\begin{array}{l@{\;\;\;}ccccccccc}
		1. &
			\pc(j) = 5 & \And &
			\setMin = 1 & \And &
			\tick = 2 & \And &
			\ticket(i) = 1 & \And &
			\ticket(j) = 3 \\
		2. &
			\pc(j) = 5 & \And &
			\setMin = 1 & \And &
			\tick = 2 & \And &
			\ticket(i) = 1 & \And &
			\ticket(j) = 1 \\
	\end{array}
\]
The decision procedure builds models that show that the VC is not
valid.  Hence, \mutex is not inductive. The formula $\mutex(i,j)$ does
not encode two important aspects of the program. First, if a thread is
in the critical section, then it owns the minimum announced ticket
(unlike in counter-model $1$) Second, the same ticket cannot be given
to two different threads (unlike in counter-model $2$).  Two new auxiliary support
invariants encode these facts:
\[
\begin{array}{lcl}
	\minticket(i) & \defsym &
		\Always( \Critical(i) \Impl \setMin = \ticket(i)) \\
	\notsame(i,j) & \defsym &
		\Always (i \neq j \land \Active(i) \land \Active(j)
			\Impl \ticket(i) \neq \ticket(j))
	\end{array}
\]
Now, \mutex can be verified using \spINV with \minticket and \notsame
as support. Unfortunately, \minticket is not inductive. The solver
reports that if two different threads $i$ and $j$ are in the critical
section with the same ticket and $\tau_6^{(j)}$ is taken, then
$\minticket(i)$ does not hold any longer. Hence, we need \notsame as
support for \minticket. However, \notsame in not inductive either. In
this case, the offending transition is $\tau_{3}$ when an existing
ticket is reused. The following invariant precludes that case:
\[
	\begin{array}{lcl}
		\activelow(i) & \defsym &
			\Always(\Active(i) \Impl \ticket(i) < \tick) \\
	\end{array}
\]
The candidate \activelow is inductive (provable using \pINV) and supports
\notsame.

\myparagraph{Mutual Exclusion for \normalfont{\CriticalSectName}:}

We proceed in a similar way. The invariants  \setmutex,
\setnotsame and \setactivelow are  identically to
\mutex, \notsame and \activelow, but \setminticket is defined as follows:
\[
	\begin{array}{lcl}
	\setminticket(i) & \defsym &
		\Always( \Critical(i) \Impl \bag.\setMin = \ticket(i)) \\
	\end{array}
\]
Similarly, \setminticket and \setnotsame support \setmutex, but
this time, \setminticket requires \setactivelow in addition to
\setnotsame as support. The extra support is needed to encode that a thread 
taking transition $\tau_3$ adds to \bag a value strictly greater than any other 
previously assigned ticket. Finally, \setnotsame relies on \setactivelow, which 
again, is inductive.

 \begin{figure}[t]
 	\centering
 	\begin{tabular}{c>{\hspace{1em}}c}
 		 \includegraphics[scale=0.4]{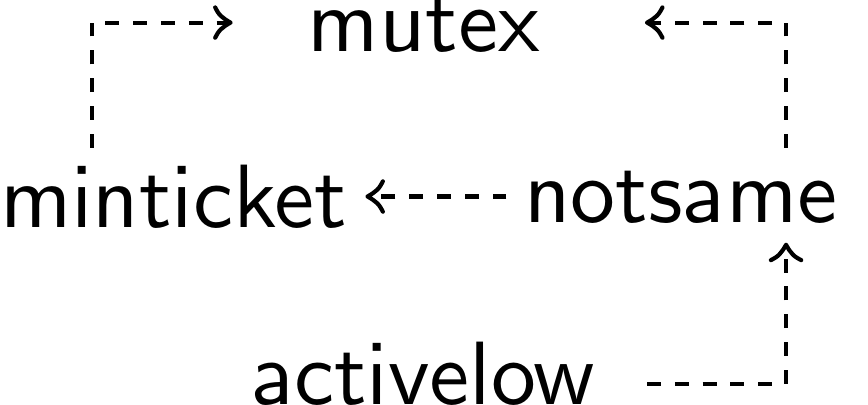} &
 		 \includegraphics[scale=0.4]{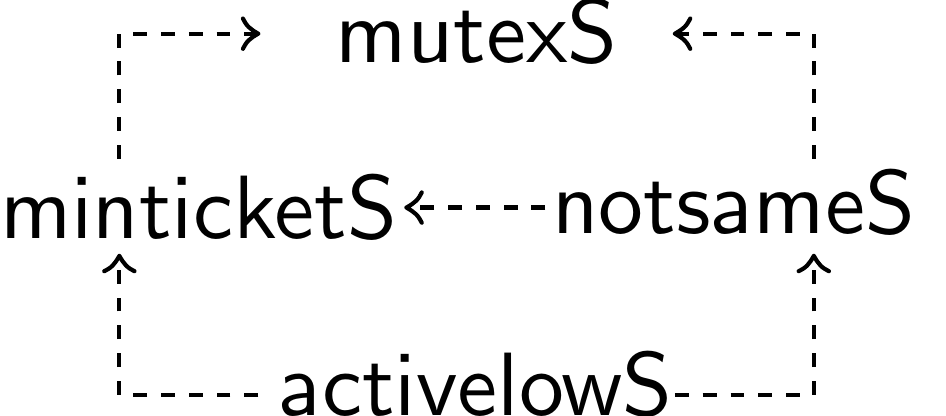} \\
 		 (a) \CriticalIntSectName &
 		 (b) \CriticalSectName \\
	\end{tabular}
 	\caption{Proof graph showing the dependencies between invariants}
 	\label{fig:proof-graphs}
 \end{figure}

 Fig~\ref{fig:proof-graphs} shows the proof graphs used for the
 empirical evaluation reported in Fig.~\ref{fig:results} in
 Section~\ref{sec:example}.


\section{Fully Symmetric Parallelism}

\newcommand{\SemanticMapsBody}{
\[\begin{array}{|c|c|}\hline
\begin{array}{rcl}
    \deltaijM(e:\sTID^\M) &=& \begin{cases}
      \begin{array}{l@{\hspace{2em}}l}
      i & \text{if $e = j$} \\[-0.5em]
      j & \text{if $e = i$} \\[-0.5em]
      e & \text{if $e \neq i,j$}
      \end{array}
    \end{cases} \\ 
    \deltaijM(e:t^\M) &=& \deltaij^t(e)\;\;\text{if
      $t\neq\sTID$} \\
    \deltaijM(f^\M) &=& \lambda x.\deltaijM(f^\M (\deltajiM x)) \\
    \deltaijM(P^\M) &=& \{ (\deltaijM x_1,\ldots,\deltaijM x_k) \;|\;\\
    & & \;\;\;\; P^\M(x_1,\ldots,x_k)\}
\end{array} &
\begin{array}{rcl}
  \deltaijE(k:\sTID) &=& \begin{cases}
    \begin{array}{l@{\hspace{2em}}l}
      i & k=j \\[-0.5em]
      j & k=i \\[-0.5em]
      k & k\neq i,j
    \end{array}
  \end{cases} \\
  \deltaijE(v:\sTID)  &=& \swap(i,j,v) \\
  \deltaijE(v[k]:\sTID) &=& \swap(i,j,v[\deltaijE(k)]) \\
  \deltaijE(c:t) &=& c \\
  \deltaijE(v:t) &=& v \\
  \deltaijE(v[k]:t) &=& v[\deltaijE(k)] \\
  \deltaijE(f(t_1\ldots t_n)) &=& f(\deltaijE(t_1)\ldots
  \deltaijE(t_n)) \\
  \deltaijP(P(t_1\ldots t_n)) &=& P(\deltaijE(t_1)\ldots
  \deltaijE(t_n)) 
\end{array} \\ \hline
\begin{array}{rcl}
  \deltaijT \tau_\ell[k] &=& \begin{cases}
    \begin{array}{l@{\hspace{2em}}l}
      \tau_\ell[i] & \text{if $k=j$}\\[-0.5em]
      \tau_\ell[j] & \text{if $k=i$}\\[-0.5em]
      \tau_\ell[k] & \text{if $k\neq i,j$}
    \end{array}
  \end{cases}
\end{array}
&
\begin{array}{rcl}
  \deltaijS(s)(v) &=& \deltaijM(s(\deltaji^E(v)))
\end{array}
\\ \hline
\end{array}
\]
}

Even though the parametrized rules \pINV and \spINV are sound for all
parametrized systems, these rules are particularly useful for
symmetric systems.
Intuitively, a parametrized transition system $\Sys[M]$ is symmetric whenever
the roles of thread ids are interchangeable, in the sense that
swapping two thread ids in a given run produces another legal run that
satisfies the corresponding temporal properties (with the ids swapped
in the property as well).
This notion of symmetry is semantic, but there are simple syntactic 
characteristics of programs that immediately guarantee symmetry. For example, if 
the only comparisons between thread identifiers in the program are for equality
and inequality, then the system is fully symmetric.
In this section, we introduce a semantic notion of symmetry and identify 
syntactic restrictions on programs that guarantee this notion of symmetry.

We show now some basic properties of fully symmetric systems.  The
essential semantic element to capture symmetry is a function
$\deltaij^t$ for each sort $t$, that defines the effect in elements of
$t$ of swapping threads $i$ and $j$. For most of the sorts, like
$\sInt$, $\sBool$ and $\Loc$ this function is simply the identity,
because thread identifiers do not interfere with values of these
types.  For $\sTID$, $\deltaij^{\sTID}$ is:
\[
\deltaij^{\sTID}(e)=\begin{cases}
  \begin{array}{l@{\;\;\;\;\;}l}
    i & \text{if $e=j$}\\[-0.5em]
    j & \text{if $e=i$}\\[-0.5em]
    e & \text{otherwise}
  \end{array}
  \end{cases}
\]
For sorts that involve thread identifiers (if present in the program),
like containers, sets, registers, etc storing elements of sort
\sTID one can easily define these semantic maps.  

Then, to characterize the effect in a run of a system of swapping two
thread ids, we define the following maps:
\begin{compactitem}
\item a model transformation map $\deltaijM$, which given a
  first-order model of the theories involved, characterizes the
  transformed model over the same domains.
\item a syntax transformation map $\deltaijE$, that allows to
  transform terms and predicates.  For variables of sort $\sTID$, the actual
  value is assigned in a concrete interpretation, so the swap between
  the ids is delegated to the interpreted function $\swap$ added to
  the theory of thread identifiers:
\[
\swap^\M(i,j,k)=\begin{cases}
  \begin{array}{ll}
    i & \text{if $k=j$}\\[-0.5em]
    j & \text{if $k=i$}\\[-0.5em]
    k & \text{otherwise}
  \end{array}
  \end{cases}
\]
\item from $\deltaijM$ and $\deltaijE$, we define the state
  transformation $\deltaijS$, that gives the program state obtained by
  swapping thread identifiers. Essentially, the valuation given to a
  transformed variable is the transformation of the value given to the
  original variable.
\item finally, $\deltaijT$ that allows to obtain the
  transition identifier that corresponds to a given transition when
  the roles of two threads are exchanged.
\end{compactitem}

%


Formally, the semantic maps $\deltaijM$, $\deltaijE$, $\deltaijS$ and
$\deltaijT$ are

\SemanticMapsBody

The essential building block used to define these transformation maps
is a swapping function $\deltaij^t$ for each sort $t$, that maps
elements in a model of the sort $t$ to the transformed elements in
$t$. This function characterizes the effect that swapping $i$ and $j$
has on elements of $t$. For most of the sorts, like $\sInt$, $\sBool$
and $\Loc$ this function is simply the identity, because thread
identifiers are not related to values of these types. For sorts that
involve thread identifiers, like set of threads $\setTh$, for example,
one can define:

\[ \deltaij^{\setTh}(S) = \left( S \setminus \{ i,j \}\right) \cup
	\left( S \setminus \{ i,j \}\right) \cup
 \left( \{ j \;|\; \textrm{if $i\in S$}\} \cup
	 \{ i \;|\; \textrm{if $j\in S$}\}\right)
\]

Similar transformations can easily be defined for containers,
registers, etc containing elements of sort \sTID. To guarantee full
symmetry all basic transformations $\deltaij^{t}$ must satisfy:
\begin{equation}
  \deltaji^{t}\circ\deltaij^{t}=\textit{id}^{\,t}.
  \label{eq:deltainvertible}
\end{equation}

\begin{figure}[t!]
	\spINVRule
  \caption{The parametrized strengthening invariance rule \spINV}
  \label{fig:spinv}
\end{figure}

\noindent From $(\ref{eq:deltainvertible})$ it follows that $\deltaijM$
satisfies $\deltaijM\circ\deltaji^\M=\deltaji^\M\circ\deltaijM=id^\M$.

For local program variables, the index is known (it is part
of the variable name), so the transformation gives the name of the
transformed variable. However, for variables of sort $\sTID$, the
actual value is assigned in a concrete interpretation, so the swap
between the ids is delegated to the interpreted function $\swap$ added
to the theory of thread identifiers.  Note that for every first-order
signature $\deltaijE$ is uniquely determined.  The following
commutativity condition is a health condition on the transformation
functions $\deltaijM$ and $\deltaijE$, where $\sem{.}$ is an
interpretation map (that gives a model in the appropriate domain to
each term and a truth value to each predicate):
\begin{equation}
  \sem{\deltaijE t} = \deltaijM \sem{t} \hspace{4em}
  \sem{\deltaijP P} \equiv \deltaijM{\sem P} 
  \label{eq:sym-transf}
\end{equation}
This condition ensures that the interpretation obtained after
transforming expression $e$, corresponds to the model transformation
of the interpretation of $e$.
Finally, note in the definition of $\deltaijS$ that first $\deltaji^E$
exchanges $v[i]$ into $v[j]$, then the interpretation $s$ is used, and
finally then the result is transformed according to the model
transformation function $\deltaijM$.

Now we are ready to define the condition for a system to be symmetric. 
\begin{definition}[Fully-Symmetric System] A parametrized system $\Sys$ is
  fully-symmetric whenever for all $M$, and for all $i,j\in [M]$, the
  following hold for all states $s$ and $s'$, transition $\tau$ and
  predicate $P$:
  \begin{compactenum}
    \item $s\vDash\Theta$ if and only if $(\deltaijS s)\vDash\Theta$.
      \label{def:sym-initiation}
    \item $\tau(s,s')$ holds if and only if $(\deltaijT{\tau})(\deltaijS
      s,\deltaijS s')$ holds.
      \label{def:sym-consecution}
    \item  $s\vDash P$ if and only
      if $(\deltaijS s)\vDash(\deltaijP P)$.
      \label{def:sym-preds}
  \end{compactenum}
  \label{def:sym}
\end{definition}

Full symmetry allows to reason about a particular thread, and conclude
the properties for arbitrary threads.


\begin{lemma}
  Let $\Sys$ be a fully symmetric system, $\varphi(\overline{k})$ be a
  parametrized formula with free variables $\overline{k}:\{k_0\ldots
  k_n\}$, and $N$ an arbitrary size.
  \[ 
  \Sys\vDash\Always\varphi(\overline{k})
  \hspace{1em}\Leftrightarrow\hspace{1em}
  \Sys[N]\vDash\bigwedge_{i_0,\ldots,i_n\in[n]}\big(\Always\varphi[i_0,\ldots,i_n]\big) 
  \]
\end{lemma}
It is important to note that the range of the concrete indices is
$[n]$, independent of the number of running threads $N$.
\begin{corollary}
  For every fully symmetric $S$ and property $\varphi(k)$
  \[ S\vDash\Always\varphi(k) 
  \hspace{1em}\Leftrightarrow\hspace{1em}
  \text{for every $N$},\; S[N]\vDash\Always\varphi[0] \]
\end{corollary}

The previous results justify the version of the strengthening
invariance rule \spINV in Fig.~\ref{fig:spinv}, where
$\Arr(\overline{k},\overline{j})$ is the set of substitutions of the
form $\sigma:\overline{k}\Into\overline{j}$.

%
Finally, for fully symmetric systems:
\begin{theorem}[Concretization]
  Let $\varphi(\overline{k})$ be with $|\overline{k}|=n$. Then:
  \[ 
  \varphi(\overline{k}) \;\; \text{is valid} 
  \;\;\; \Leftrightarrow\;\;\; 
  \bigwedge\limits_{\alpha\in A} \alpha(\varphi)\;\;\text{is valid} 
  \]
  where $A=\Arr(\overline{k},[n])$ is the set of concretizations of
  variables in $\Var(\varphi)$.
\end{theorem}

For example, if one intends to prove that $p(i)$ is inductive,
the concretization theorem allows to reduce \Premise{P3} in \pINV to:
\begin{equation} 
  p[0] \And \tau[1] \Into p'[0] 
  \label{eq:concex1}
\end{equation}
where $p[0]$ is a short for $\alpha(p(i))$ with $\alpha:i\Into 0$.
Formula (\ref{eq:concex1}) involves no arrays. Similarly, to show
$\Always p(i)$ with support invariant $q(j)$, rule S5 can be reduced
to:
\[
\begin{array}{rcl}
   q[0] \And q[1]  \And p[0] \And \tau[1] &\Into& p'[0]\\
\end{array}
\]

In practice, the concretization is not performed upfront before
discharging the verification condition to the SMT-Solver. Our use of
arrays to encode parametrized formulas can be handled using the theory
of uninterpreted functions and let the solver perform the search and
propagation.




\section{Detailed Invariants for Case Studies} 
\label{app:implementation}

We prove that the program in Fig~\ref{fig:listcode} satisfies:
\begin{inparaenum}[(1)]
\item list shape preservation; and
\item the list implements a set, whose elements correspond to those
  stored in \elements.
\end{inparaenum}
The theory \TLLThree allows to reason about addresses, elements,
locks, sets, order, cells (\ie list nodes), memory and reachability.
A cell is a struct containing an element, a pointer to next node in
the list and lock to protect the cell. A lock is associated with
operations $\lock$ and $\unlock$ to acquire and release. The memory
(\heap) is modeled as an array of cells indexed by addresses.
The specification is:
\[
	\formulaFullListReduced
\]
Formula \phiPreserve is 0-index since it only constrains global
variables. (L1) establishes that \fNull belongs to \region and that
\region is exactly the set of addresses reachable in the \heap
starting from \head, which ensures that the list is acyclic. (L2) and
(L3) express some sanity properties of the sentinel nodes \head and
\tail.  Finally, (L4) establishes that \elements is the set of
elements in cells referenced by addresses in \region, and that the
list is ordered. The main specification is \invPreserve, defined as
$\Always \phiPreserve$.

\Leap can establish that \invPreserve holds initially, but fails to
prove that \invPreserve is preserved by all transitions (\ie
\invPreserve is not a parametrized invariant), so support invariants
are required.
To prove (L1) the support invariant \phiRegion captures how addresses
are added and removed from \region in the program. Local variable $v$
in procedure \MostGeneralClient, \Search, \Insert and \Remove, is
denoted by \local{C}{v}, \local{S}{v}, \local{I}{v} and \local{R}{v}
respectively:
\[
	\formulaFullRegionReduced
\]
Formula \phiRegion is $1$-index and determines which addresses belong
to \region depending on each program location. Invariant
$\invRegion(i)$ is defined as $\Always \phiRegion(i)$. 
Invariant $\invNext(i)$ captures the relative position in the list of
the cells pointed by \head and \tail and local variables \prev, \curr
and \aux. The details of can be found in the appendix.  Invariant
$\invNext(i)$ is needed for (L2).
To prove (L3) and (L4) we need to show that order is preserved.  We
express this constraint with formula \phiOrder:
\[
	\formulaFullOrderReduced
\]
\noindent and define invariant $\invOrder(i)$ as $\Always
\phiOrder(i)$.  Invariant \invLock captures those program locations at
which a thread owns a cell in the heap:
\[
	\formulaFullLockReduced
\]
Finally, formula $\phiDisjoint$ encodes that two different
threads calls to \emph{malloc} return two different addresses:
\[
	\formulaFullDisjoint
\]
In this case, $\invDisjoint (i,j)$, defined as $\Always \phiDisjoint(i,j)$ 
is a 2-index invariant.

In practice, when proving concurrent datatypes these candidate
invariants are easily spotted using the information obtained from an
unsuccessfully attempt of \Leap to prove a particular VC. \Leap
parses the counterexample (model) returned by the SMT solver, which is
usually very small, involves few threads and allows to easily
understand the missing intermediate facts.
%
%
Fig.\ref{fig:proof-graph} shows the proof graph for the verification
of concurrent lock-coupling lists. In the graphs, a dashed arrow from
$\varphi$ to $\psi$ denotes that $\varphi$ is used as support for
$\psi$.
\Leap parses proof graphs as input and applies \gINV when necessary.
Additionally, graphs can specify program locations for which to apply
a particular formula as support, which greatly speeds proof checks.
Fig.~\ref{fig:results} shows the results of this empirical evaluation,
executed on a computer with a 2.8 GHz processor and 8GB memory.  The
columns show the index of the formula; the total number of generated
VCs; the number of VCs verified using the position based DP; the
number of VCs verified using the specialized DP and, finally, the
total running time in seconds required to verify all VCs. We benchmark
the times in four different scenarios using different tactics. The
first scenario (FS) uses \spINV with full support, that is, all
invariant candidates are used as support. The second scenario (FA)
considers only full assignments when generating support.  The third
scenario (FA-SS) involves full assignments in addition to discarding
superfluous support information. The last column reflects the forth
scenario, using proof graphs. We use OM to represent
out-of-memory failure.  These results show that, in practice, tactics
are essential for efficiency when handling non-trivial examples such
as concurrent lists.

\section{Proof Graph for Concurrent Lock-coupling Lists}

We present now the full proof graph for this concurrent
lock-coupling single-linked lists implementation. We use the following
notation to represent a proof-graph:

\begin{verbatim}
-> inv [l1:P:sup1, sup2, sup3;
        l2:P:sup4, sup5
       ] { SMP : pretactic | posttactic }
\end{verbatim}

\noindent where \verb+inv+ is the invariant candidate to be verified. Next, 
between brackets it is possible to specify invariant support for specific 
program locations. This argument is optional. Required support invariants are 
provided as a list of support rules, separated by \verb+;+. Each support rule 
consists on a location, a possible premise identifier to localize the support 
generation on a specific invariant rule premise and a list of invariants to be 
used as support. Finally, it is possible to describe the method used to compute 
the domain bounds for the small model property, as well as tactics to be used in 
support generation and formula simplification, separated by a \verb+|+.

Fig.\ref{fig:proofgraph-lists} shows the proof graph for the current 
lock-coupling list example.

\begin{figure}[p]
\begin{verbatim}
-> fullOrder {pruning:reduce2|simpl}

-> fullPreserve [34:E:fullRegion;
                 35:E:fullRegion,fullNext,fullLock,fullOrder 
                        {pruning:reduce2|simpl};
                 51:E:fullNext,fullRegion,fullOrder
                        {pruning:reduce2|simpl}
                ] {pruning:reduce2|simpl}

-> fullNext [ 5:N:fullPreserve;
             30:N:fullPreserve,fullRegion;
             34:N:fullRegion;
             34:E:fullRegion,fullDisjoint;
             35:N:fullRegion;
             35:E:fullLock,fullRegion;
             41:N:fullPreserve,fullRegion;
             47:N:fullPreserve,fullRegion;
             50:N:fullPreserve,fullRegion;
             51:N:fullPreserve,fullRegion;
             51:E:fullPreserve,fullLock,fullRegion
            ] {pruning:reduce2|simpl}

-> fullLock [28:N:fullNext;
             29:N:fullNext;
             36:N:fullNext;
             45:N:fullNext;
             46:N:fullNext;
             52:N:fullNext
            ] {pruning:reduce2|simpl}

-> fullDisjoint {pruning:reduce2|simpl}

-> fullRegion [24:N:fullPreserve {pruning:reduce2|simpl};
               28:N:fullNext;
               30:N:fullPreserve;
               35:E:fullDisjoint;
               41:N:fullPreserve;
               45:N:fullNext;
               47:N:fullPreserve,fullNext;
               51:N:fullPreserve,fullNext;
               51:E:fullPreserve,fullNext,fullLock
                        {pruning:reduce2|simpl}
              ] {pruning:reduce2|simpl}
\end{verbatim}
\caption{Proof graph representation for concurrent lock-coupling lists}
\label{fig:proofgraph-lists}
\end{figure}


\end{document}